\documentclass[9pt,twocolumn,twoside,lineno]{pnas-new}

\templatetype{pnasresearcharticle} 

\usepackage{amsmath,amssymb}

\usepackage{graphicx}
\usepackage{dcolumn}
\usepackage{bm}
\usepackage{hyperref}

\graphicspath{ {figs/} }

\usepackage{mathtools}
\usepackage{amsmath}
\usepackage{amsbsy}
\usepackage{amssymb}
\usepackage{amscd}
\usepackage{amsfonts}
\usepackage{bm}
\usepackage{graphics}
\usepackage[plain]{fancyref}
\usepackage{xcolor}
\usepackage{mdframed,color}
\usepackage{graphicx}

\newcommand{\eff}{\text{eff}}

\usepackage{hyperref}
\hypersetup{backref, colorlinks=true}
\usepackage{comment}

\DeclareMathOperator{\csch}{csch}

\newcommand{\spectralRadius}[1]{\kappa\left(#1\right)}
\newcommand{\rmag}{r} 
\newcommand{\vvec}[1]{\bm{#1}}
\newcommand{\rvec}{\vvec{\rmag}}
\newcommand{\dir}[1]{\hat{#1}}
\newcommand{\rdir}{\dir{\rvec}}
\newcommand{\tens}[1]{\bm{#1}}

\newcommand{\mLen}{b} 
\newcommand{\n}{n} 
\newcommand{\chainStretch}{\gamma} 
\newcommand{\chainDensity}{N} 
\newcommand{\ChainDensity}{N_0} 
\newcommand{\popNum}{m} 
\newcommand{\numPatches}{\mathcal{N}}
\newcommand{\nvec}{\hat{\vvec{n}}} 
\newcommand{\partitionFunction}{\mathcal{Z}} 
\newcommand{\nDensity}{\rho} 
\newcommand{\nC}{C} 
\newcommand{\laMult}{\tau} 
\newcommand{\laMultVec}{\vvec{\laMult}} 

\newcommand{\eFieldMag}{E} 
\newcommand{\eFieldVec}{\vvec{\eFieldMag}}
\newcommand{\dipoleMag}{\mu} 
\newcommand{\dipoleVec}{\vvec{\dipoleMag}}
\newcommand{\nvecToDipoleij}{M} 
\newcommand{\nvecToDipole}{\tens{\nvecToDipoleij}}
\newcommand{\eTransform}{\mathcal{E}}
\newcommand{\eTransformVec}{\vvec{\eTransform}}

\newcommand{\mEnergy}{u} 
\newcommand{\trans}[1]{#1^{T}} 
\newcommand{\collection}[1]{\left\{#1\right\}} 

\newcommand{\invT}{\beta} 
\newcommand{\cdGibbsChain}{\mathcal{F}} 

\newcommand{\unitSphere}{\mathbb{S}^2}
\newcommand{\df}[1]{\text{d}#1} 
\newcommand{\intOverSphere}[1]{\int_{\unitSphere} \df{A} \: #1 } 
\newcommand{\avgOverM}[1]{\intOverSphere{\nDensity #1}} 
\newcommand{\nMult}{\xi}
\newcommand{\nMultVec}{\vvec{\nMult}} 
\newcommand{\Lang}{\mathcal{L}}
\newcommand{\invLang}{\Lang^{-1}}

\newcommand{\chainPolarMag}{p} 
\newcommand{\chainPolarVec}{\vvec{p}} 

\newcommand{\pullback}[1]{\tilde{#1}}
\newcommand{\Rmag}{\pullback{\rmag}}
\newcommand{\Rvec}{\pullback{\rvec}}
\newcommand{\avg}[1]{\left< #1 \right>}
\newcommand{\avgOverR}[1]{\avg{#1}_{\Rvec}}
\newcommand{\avgOverr}[1]{\avg{#1}_{\rvec}}
\newcommand{\generic}{\Box}
\newcommand{\cdGibbsDensity}{\mathcal{W}^*}
\newcommand{\polarMag}{P} 
\newcommand{\polarVec}{\vvec{\polarMag}} 
\newcommand{\HelmholtzDensity}{\mathcal{W}}

\newcommand{\F}{\tens{F}}
\newcommand{\J}{J}
\newcommand{\PolarMag}{\pullback{\polarMag}}
\newcommand{\PolarVec}{\pullback{\polarVec}}

\DeclareMathOperator{\divv}{div}

\newcommand{\diverge}[1]{\divv #1}

\newcommand{\charge}{\varrho}
\newcommand{\chargeFree}{\charge_f}
\newcommand{\vacPerm}{\epsilon_0}

\newcommand{\Gradd}{\mathrm{Grad}}

\newcommand{\takeGrad}[1]{\Gradd \: #1}

\newcommand{\xj}{x}
\newcommand{\yj}{y}
\newcommand{\xvec}{\vvec{\xj}} 
\newcommand{\yvec}{\vvec{\yj}} 
\DeclareMathOperator{\diag}{diag} 
\newcommand{\pStretchSymbol}{\lambda} 
\newcommand{\pStretch}[1]{\pStretchSymbol_{#1}} 
\newcommand{\pZero}{\overline{\pStretchSymbol}}
\newcommand{\dpStretch}[2]{\pStretch{#1,#2}} 
\newcommand{\identity}{\tens{I}}
\newcommand{\euclid}[1]{\hat{\vvec{e}}_{#1}}

\newcommand{\orderOf}[1]{\mathcal{O}\left(#1\right)}
\DeclareMathOperator{\Size}{Size}

\newcommand{\smallpar}{\epsilon}
\newcommand{\cellLength}{\ell}

\title{Flexoelectricity in soft elastomers and the molecular mechanisms underpinning the design and emergence of giant flexoelectricity}

\author[a,b,1]{Matthew Grasinger}
\author[c,1]{Kosar Mozaffari} 
\author[c,d,2]{Pradeep Sharma}

\affil[a]{Materials and Manufacturing Directorate, Air Force Research Laboratory, WPAFB, OH 45433, USA}
\affil[b]{UES Inc., Dayton, OH 45432, USA}
\affil[c]{Department of Mechanical Engineering, University of Houston, Houston, TX 77204, USA}
\affil[d]{Department of Physics, University of Houston, Houston, TX 77204, USA}

\leadauthor{Sharma} 

\significancestatement{The molecular-scale phenomena that lead to the emergence of the flexoelectric effect in elastomers are not well understood. We present a molecular-to-continuum scale theory and show it to be valid for common types of elastomers. We unravel a mechanism for achieving giant flexoelectricity--which finds support in prior experimental results. The multiscale theory is then leveraged for material design and we propose specific modalities for (1) the emergence of piezoelectricity, (2) tuning the direction of flexoelectricity, and (3) flexoelectricity which is invariant with respect to spurious deformations. These developments may pave the way for the future development of efficient soft energy harvesters, sensors and actuators, and may prove useful for high degree of freedom soft robots and artificial muscles.}

\equalauthors{\textsuperscript{1}Authors M. G. and K.M. contributed equally to this work.}
\correspondingauthor{\textsuperscript{2}To whom correspondence should be addressed. E-mail: psharma@uh.edu}


\begin{abstract}
Soft robotics requires materials that are capable of large deformation and amenable to actuation with external stimuli such as electric fields. Energy harvesting, biomedical devices, flexible electronics and sensors are some other applications enabled by electro-active soft materials. The phenomenon of flexoelectricity is an enticing alternative that refers to the development of electric polarization in dielectrics when subjected to strain gradients. In particular, flexoelectricity offers a direct linear coupling between a highly desirable deformation mode (flexure) and electric stimulus. Unfortunately, barring some exceptions, the flexoelectric effect is quite weak and rather substantial bending curvatures are required for an appreciable electro-mechanical response. Most experiments in the literature appear to confirm modest flexoelectricity in polymers although perplexingly, a singular work has measured a ``giant'' effect in elastomers under some specific conditions. Due to the lack of an understanding of the microscopic underpinnings of flexoelectricity in elastomers and a commensurate theory, it is not currently possible to either explain the contradictory experimental results on elastomers or pursue avenues for possible design of large flexoelectricity. In this work, we present a statistical-mechanics theory for the emergent flexoelectricity of elastomers consisting of polar monomers. The theory is shown to be valid in broad generality and leads to key insights regarding both giant flexoelectricity and material design. In particular, the theory shows that, in standard elastomer networks, combining stretching and bending is a mechanism for obtaining giant flexoelectricity, which also explains the aforementioned, surprising experimental discovery.
\end{abstract}

\doi{\url{https://doi.org/10.1073/pnas.2102477118}}

\begin{document}

\verticaladjustment{-2pt}

\maketitle
\thispagestyle{firststyle}
\ifthenelse{\boolean{shortarticle}}{\ifthenelse{\boolean{singlecolumn}}{\abscontentformatted}{\abscontent}}{}

\section{Introduction}

\begin{figure}[hbt!]
\begin{center}
\includegraphics[width=0.95\linewidth]{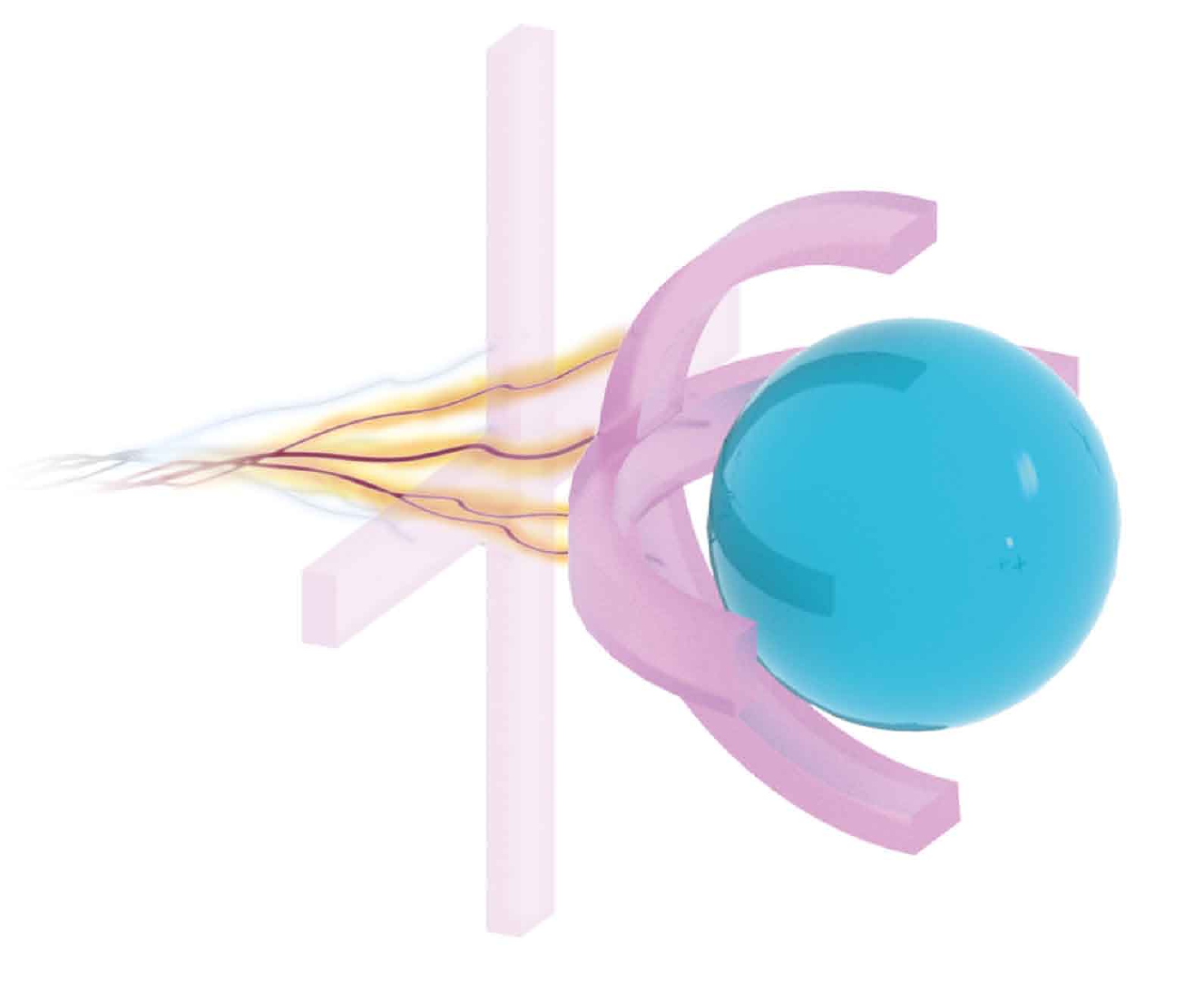}
\caption{\label{fig:ref-fig1} Schematic of a soft robotic appendage, deforming in order to grab an object due to electrical stimuli.}
\end{center}
\end{figure}

\dropcap{T}ypical \emph{hard} materials such as ceramics have elastic moduli in the order of several hundreds of giga-Pascals. Their deformation, under appropriate levels of mechanical forces, is barely discernible to the naked eye. On the other end of the spectrum, a change of area of nearly 1700$\%$ has been demonstrated for an acrylic membrane \cite{keplinger2012harnessing}. While both \emph{hard} and \emph{soft} materials have their respective usage, there are important technological imperatives to consider highly deformable soft materials. An oft-quoted example is that of a robotic appendage clasping and moving an object (Fig~\ref{fig:ref-fig1}). This requires the capability of large deformation. Ideally, we also hope that such a motion is accomplished through the application of a stimulus that is easily applied (e.g. an electrical field through a battery) and with a modest expenditure of energy. Soft materials that are highly deformable, tough \cite{sun2012highly}, and respond easily to stimuli such as electric field (the focus of this work) have wide applications: biomedical prostheses and devices \cite{kwak2020wireless,rogers2010materials,zhang2020water,zhao2019mechanics,ze2020magnetic}, shape-conforming sensors \cite{xu2014soft,kang2016bioresorbable,kim2012thin,kim2011epidermal}, energy harvesting \cite{han2019three,erturk2011piezoelectric,nan2018compliant}, stretchable electronics \cite{kim2009large,khang2006stretchable,xu2013stretchable,kim2018printing}, and of course robotics \cite{han2019three,kim2007electroactive,kim2019ferromagnetic}. As a compelling illustration of applications, we refer the reader to fascinating recent works on using magnetic actuation to achieve a soft robot made essentially of a rubber filament that is capable of being steered through the complex cerebrovascular system with aneurysms \cite{kim2019ferromagnetic}.\\

The electromechanical coupling that we \emph{hope} for a soft material to possess is \emph{piezoelectricity}. Materials that are piezoelectric permit a direct and a linear relation between the development of an electric field when subjected to a mechanical stress, and vice-versa. Piezoelectricity can be shown by a linearized relation between polarization($\bold{P}$) and the strain tensor($\pmb{\varepsilon}$) as ${P}_i\sim\mathcal{D}_{ijk}{\varepsilon}_{jk}$ where $\pmb{\mathcal{D}}$ is a third order tensor reflecting the piezoelectric properties of the material. Unfortunately, piezoelectricity requires the atomic structure of the material to conform to a rather stringent set of symmetry conditions found only in certain hard ceramics. The rather few piezoelectric polymers such as PVDF are (relatively speaking) fairly stiff\footnote{Their elastic modulus is of the order of a few GPa's while the soft materials that are highly deformable often need to have stiffness in the KPa range}. Therefore, the electromechanical coupling mechanism that we rely on in the context of soft dielectric elastomer is the Maxwell stress effect or electrostriction---indeed these are the precise mechanisms operative in the example we quoted in the preceding paragraph involving the nearly 1700$\%$ areal increase in the acrylic membrane \cite{keplinger2012harnessing}. Both the Maxwell stress effect and electrostriction are universally present in all dielectrics where the induced mechanical strain $\pmb{\varepsilon}$ scales with the square of the applied electric field $E^2$. There is no converse effect hence applications such as energy harvesting and sensing are not easily possible. Further, rather high electric fields must be imposed to generate sufficient forces for an appreciable actuation. An exciting alternative is to embed frozen charges and dipoles in soft matter to create so-called electrets \cite{sessler1999electromechanical,sessler1980physical,wegener2005microstorms,buchberger2008flexible,deng2014electrets, liu2018emergent}.
 Through the interaction of pre-existing immobile charges, electrostriction, and the inherent softness of the material, electrets effectively act as piezoelectrics{\footnote{Detailed theoretical framework outlining the conditions for when electrets act a piezoelectric or even pyroelectric are available here \cite{liu2018emergent,darbaniyan2019designing}. Other illustrative examples may be found in the literature \cite{darbaniyan2021soft,deng2014electrets,apte20202d}.}}. While this is a promising research avenue, a key impediment in the use and development of electret materials is that the frozen external charges are only meta-stable and have a propensity to leak out--especially at elevated temperatures and under humid conditions \cite{mellinger2006thermal}. \\

Having established the relevant context regarding electromechanical coupling mechanisms in soft materials, we now turn to a yet another phenomenon--\emph{flexoelectricity}--which has generated much recent attention \cite{krichen2016flexoelectricity,deng2017continuum,nguyen2013nanoscale}. Briefly, it is the coupling between polarization and strain gradients and can be illustrated by ${P}_i\sim f_{ijkl} \left({\partial \varepsilon_{jk}}/{\partial x_l} \right)$ where $\pmb{f}$ is a fourth order tensor embodying the flexoelectric behavior of the material. Like electrostriction, flexoelectricity is also a universal phenomenon which occurs in all insulating materials. In non-piezoelectrics, the symmetry of the charge density throughout the material is such that the dipole moment per unit volume, i.e. polarization, vanishes. A homogeneous deformation cannot break the symmetry of the ground state charge density; however, strain gradients can and do, thereby inducing polarization. 
Very interestingly, the flexoelectric effect has been used to create apparently piezoelectric materials without using piezoelectric materials~\cite{sharma2010piezoelectric,fousek1999possible,chu2009flexure,nguyen2013nanoscale} and has broad-ranging applications in energy harvesting~\cite{wang2019flexoelectricity,deng2014nanoscale,jiang2013flexoelectric,majdoub2008dramatic,majdoub2009erratum,rahmati2019nonlinear,choi2017measurement}, and sensors and actuators~\cite{bhaskar2016flexoelectric,bhaskar2016bflexoelectric,wang2013giant,abdollahi2015constructive}. The phenomenon also appears to be implicated in several biophysical functions, e.g. the mammalian hearing mechanism \cite{deng2019collusion,brownell2001micro,breneman2009hair}. A notable aspect of flexoelectricity is the associated size-effect \cite{krichen2016flexoelectricity}. The developed polarization due to flexoelectricity scales with the gradient of the strain. Large strain gradients are most easily produced when feature size is small (e.g. beam thickness, size of embedded inclusions in a composite) and indeed, at least for hard materials, the phenomenon becomes appreciable only at the nanoscale. For soft matter, due to the corresponding lower elastic modulus, a non-trivial flexoelectric response is even possible at micron-length scales \cite{deng2014flexoelectricity, rahmati2019nonlinear,wen2019flexoelectret}. As a notable example, Deng et. al.~\cite{deng2014flexoelectricity} exploited flexoelectricity to design soft materials whose apparent piezoelectric strength is nearly 20 times larger than the hard ferroelectrics like barium titanate with feature sizes of the order of $0.6$ microns.\\

The phenomenological theory of flexoelectricity with various degrees of sophistication and nuances, appears to be now well-developed (e.g. \cite{maranganti2006electromechanical, deng2017continuum,codony2021modeling}). In the context of soft matter, a first attempt appeared in \cite{deng2014flexoelectricity} and we highlight a more recent paper by \cite{codony2021modeling} that is specifically targeted towards proper accounting for large deformations and the correct interpretation of the flexoelectric matter property tensor for large deformations. Significant efforts have also been expended towards computational approaches (e.g. \cite{abdollahi2014computational,yvonnet2017numerical,codony2021modeling,thai2021staggered}).  At the other end of the scale-spectrum, extensive research has also ensued on the microscopic underpinnings of the phenomenon. The atomistic-scale mechanisms for flexoelectricity are now relatively well understood for crystalline materials~\cite{tagantsev1986piezoelectricity,maranganti2009atomistic, stengel2013flexoelectricity,stengel2014surface,dreyer2018current}, liquid crystals~\cite{de1993physics,meyer1969piezoelectric}, and 2D materials~\cite{kothari2018critical,kothari2019critical,kvashnin2015flexoelectricity,ahmadpoor2015flexoelectricity,kumar2021flexoelectricity,banerjee2016cyclic}. However, a similar molecular-scale understanding of the flexoelectricity of elastomers has remained elusive~\cite{krichen2016flexoelectricity}. This is a key impediment in the design of next-generation soft materials that exhibit substantive flexoelectricity. The flexoelecric coefficients of most soft polymers (like elastomers) exhibit rather modest values. Attempts were made in~\cite{marvan1997static} and~\cite{marvan1998flexoelectric} to explain flexoelectricity in elastomers through quasi-1D microscopically-based models which utilized linear elastic chains connecting neutral, positive, and negative ions and a kind of free volume theory, respectively.
These models are useful to assess the order of magnitude of the flexoelectric effect in elastomers but can hardly capture the nuances of the actual mechanisms and thus are of limited utility for designing materials.  The state of affairs is complicated further by experiments which indicate a wide range of flexoelectricity in polymers~\cite{baskaran2012strain,chu2012flexoelectricity}. In this work, we take inspiration from the fact that statistical mechanics and polymer network modeling have explained much of what we understand about the microscopic underpinnings of elasticity~\cite{treloar1975physics,flory1944network,arruda1993threee,boyce2000constitutive} in elastomers. More recently, a similar success has also been achieved in the context of electroelasticity~\cite{cohen2016electromechanical,itskov2018electroelasticity,grasinger2020architected} which has inspired our current approach. \\

In this work: 

\begin{enumerate}
    \item We create a molecular-scale theory for the emergent flexoelectricity of elastomers with polar monomers. The molecular-scale theory is quite general; it is valid for two of the most common theoretical models for polymers: the freely jointed chain and the wormlike chain.
    \item We provide an explanation for the rather wide and non-intuitive range of flexoelectric properties measured for polymers and specifically explain the unusually large values obtained when stretching and bending are combined.
    \item We provide an interpretation of understanding microscopic flexoelectricity using the notion quadrupolar moments and leverage this connection to develop material design principles for tuning the direction of the flexoelectric effect, as well as for stretch-invariant flexoelectricity.
      \item Finally, we provide simple and clear guidelines to design soft elastomers with giant flexoelectricity.
\end{enumerate}

\section{Statistical mechanics of the electroelasticity of a freely jointed chain}

As a starting point, we derive the free energy of a polymer chain in a fixed ambient temperature, $T$, with at a fixed end-to-end vector, $\rvec$, subjected to a constant electric field, $\eFieldVec$.
We work in the constant electric field ensemble as a matter of convenience and will eventually use a Legendre transformation to arrive at the Helmholtz free energy.
To work in this ensemble, we need the variation of the electric field due to the presence of monomer dipoles to be negligible; that is, the monomers need to be approximately non-interacting\footnote{To be precise, we assume that the energy of dipole-dipole interactions is small relative to the thermal energy of the chain, $\n k T$. An order of magnitude estimate for this condition can be obtained from dimensional analysis: $\tilde{\mu}^2 \ll 1$ where $\tilde{\mu} \coloneqq \mu / \sqrt{\epsilon_0 \mLen^3 k T}$ and $\mu$ is a characteristic dipole magnitude.}.
Such an assumption is quite reasonable c.f.~\cite{grasinger2020statistical} and \S 1 of the SI where we use Markov chain Monte Carlo calculations to prove this point.\\

We assume that the energy required to deform an individual monomer is much greater than $kT$; and, hence, we model the monomers as rigid with length $\mLen$.
For the freely jointed chain, monomers are also free to rotate about their neighboring bonds.
We denote the number of monomers per chain as $\n$, the chain stretch as $\chainStretch \coloneqq \rmag / \n \mLen$, and the orientation of a monomer as $\nvec$ so that $\rvec = \mLen \sum_{i=1}^{\n} \nvec_i$.
By assumption, the monomers are polar; or, in other words, have a permanent dipole attached to them.
In general, the dipole vector, $\dipoleVec$, need not be aligned with $\nvec$.
However, we do assume that the dipole is a linear transformation of $\nvec$; i.e.
$\dipoleVec = \nvecToDipole \nvec$.
If, for example, a dipole of fixed magnitude is attached rigidly to the monomer such that it rotates with the monomer, then $\nvecToDipole = \dipoleMag \tens{Q}$ where $\tens{Q} \in \mathrm{SO}\left(3\right)$.
In any case, the electric potential of a monomer depends on its orientation.
Let $\eTransformVec \coloneqq \trans{\nvecToDipole} \eFieldVec$. 
Then the electric potential of a monomer is given by the expression:
\begin{equation}
    \mEnergy = -\dipoleVec \cdot \eFieldVec = -\left(\nvecToDipole \nvec\right) \cdot \eFieldVec = -\eTransformVec \cdot \nvec.
\end{equation}

In order to formulate the partition function, we divide the unit sphere into $\numPatches$ infinitesimal patches of area with unit normal $\nvec_i, i = 1, 2, ..., \numPatches$.
Let $\popNum_i$ denote the number of monomers in the chain with direction $\nvec_i$.
Then,
\begin{equation} \label{eq:partition-function}
    \partitionFunction = \sum_{\collection{\popNum_j}'} \left\{ \exp \left[\invT \sum_{i=1}^{\numPatches} \popNum_i \eTransformVec \cdot \nvec_i \right] \frac{\n!}{\prod_{i=1}^{\numPatches} \popNum_i !} \right\},
\end{equation}
where $\invT = 1 / k T$ and the prime over $\collection{\popNum_j}$ denotes that the summation is over all of the collections of population numbers that satisfy the constraints: $\sum_i \popNum_i = \n$ and $\mLen \sum_i \popNum_i \nvec_i = \rvec$.
Enumerating over all of the terms in \eqref{eq:partition-function} is difficult in general.
Instead, we approximate the logarithm of the sum by the logarithm of its maximum-term and, taking $\numPatches \rightarrow \infty$ (i.e. the continuum limit of the population numbers), arrive at a mean-field theory.
In doing so, we find that
\begin{equation} \label{eq:nDensity}
    \nDensity\left(\nvec\right) = \nC \exp \left[\invT \eTransformVec \cdot \nvec + \laMultVec \cdot \nvec\right],
\end{equation}
where $\nDensity$ is a probability density function associated with finding a monomer with some direction $\nvec$, and $\nC$ and $\laMult$ are Lagrange multipliers that are determined by enforcing the continuum analog of the normalization and end-to-end vector constraints, respectively:
\begin{equation}
    \label{eq:densityConstraints}
   \n = \avgOverM{ }, \quad 
   \frac{\rvec}{\mLen} = \avgOverM{\nvec}.
\end{equation}
Once the unknown multipliers have been determined, the closed-dielectric free energy, $\cdGibbsChain$, is approximately
\begin{equation}
    \label{eq:cdGibbsChain}
    \cdGibbsChain = -\frac{1}{\invT} \ln \partitionFunction \approx \avgOverM{\mEnergy} + \frac{1}{\invT}\avgOverM{\ln \nDensity} .
\end{equation}

Let $\nMultVec \coloneqq \invT \eTransformVec + \laMultVec$.
Then the solution for $\nC$ and $\nMultVec$ is well-known~\cite{kuhn1942beziehungen} and
\begin{equation}
    \label{eq:kg-solution}
    \nC = \frac{\n \nMult \csch \nMult}{4 \pi}, \quad
    \nMultVec = \invLang\left(\chainStretch\right) \rdir,
\end{equation}
where $\nMult = |\nMultVec|$, $\rdir = \rvec / \rmag$, and $\invLang$ is the inverse Langevin function.
Based on cursory examination of  \eqref{eq:kg-solution}, it may seem somewhat surprising that $\nDensity$ is independent of the electric field and is, in fact, equivalent to the $\nDensity$ that maximizes the entropy. However, it is easy to show that this is indeed the $\nDensity$ which minimizes the free energy for any monomer potential energy which is linear in $\nvec$.
To show this, let $\nDensity'$ be any function which satisfies the constraints, \eqref{eq:densityConstraints}; and, let $\mEnergy\left(\nvec\right) = \alpha + \mathbf{A} \nvec$.
\newcommand{\auxdrho}{\delta \nDensity}
Now, we can perturb $\nDensity$ with some arbitrary function $\auxdrho\left(\nvec\right)$ and still satisfy the constraints provided that
\begin{equation} \label{eq:drhoConstraints}
    \intOverSphere{\auxdrho} = 0, \quad
    \intOverSphere{\left(\auxdrho\right) \nvec} = 0.
\end{equation}
Then the change in internal energy due to $\nDensity \rightarrow \nDensity + \auxdrho$ is
\begin{equation*}
    \intOverSphere{\left(\nDensity+\auxdrho\right)\mEnergy}-\avgOverM{\mEnergy} = \intOverSphere{\left(\auxdrho\right)\left(\alpha + \mathbf{A} \nvec\right)}
\end{equation*}
which vanishes by \eqref{eq:drhoConstraints}.
Therefore, if the monomer potential energy is linear in $\nvec$, then all $\nDensity$ which satisfy \eqref{eq:densityConstraints} have the same internal energy.
In conclusion, we infer that, in this case, the density which minimizes the free energy is the same as that which maximizes the entropy. \\

Next, using \eqref{eq:kg-solution} and \eqref{eq:nDensity} in \eqref{eq:cdGibbsChain}, we obtain the chain free energy:
\begin{equation} \label{eq:cdGibbsChain-solution}
      \cdGibbsChain = \n \left[\frac{1}{\invT} \left(\chainStretch \nMult + \ln \frac{\nMult \csch \nMult}{4 \pi}\right) - \chainStretch\left(\eTransformVec \cdot \rdir\right)\right].
\end{equation}
Finally, we define the chain polarization as $\chainPolarVec = \avgOverM{\dipoleVec}$.
It is easy to show that $-\partial \cdGibbsChain / \partial \eFieldVec = \chainPolarVec$
\footnote{In general, the relationship between $\eFieldVec$ and $\chainPolarVec$ requires solving a nonlocal boundary value problem.
However, when dipole-dipole interactions are negligible, the nonlocal relationship simplifies to the local one given here~\cite{grasinger2020statistical}.
}.
Thus,
\begin{equation}\label{eq:plrztn-1st-apprch}
    \chainPolarVec = \nvecToDipole \frac{\rvec}{\mLen} = \n \chainStretch \nvecToDipole \rdir,
\end{equation}
where $\rdir = \rvec / \rmag$.
By \eqref{eq:plrztn-1st-apprch}, the chain polarization is determined uniquely by the chain end-to-end vector, irrespective of the temperature or the electric field. 
In addition, since $\chainStretch \in \left[0, 1\right]$ we have that $\chainPolarMag \in \left[0, \n \spectralRadius{\nvecToDipole}\right]$ where $\spectralRadius{\generic}$ is the spectral radius of $\generic$.
If $\nvecToDipole = \dipoleMag \tens{Q}$ then the maximum polarization simplifies to $\n \dipoleMag$ and occurs when the chain is fully stretched.
Importantly, despite the fact that dipole-dipole interactions were assumed negligible in deriving \eqref{eq:plrztn-1st-apprch}, we note that Markov chain Monte Carlo simulations agree with \eqref{eq:plrztn-1st-apprch} nearly exactly, even when dipole-dipole interactions \emph{are} significant (see \S1 of the SI).

\section{Statistical mechanics of the electroelasticity of a wormlike chain}

Ultimately, because we are interested in what universal phenomena may arise from all polymer chains with polar monomers--not just those that behave as freely jointed chains--we next consider polar monomers in a wormlike chain.
By wormlike chain, we mean a chain which again consists of rigid monomers but has some bending stiffness along its contour.
That is, the bond which joins monomer $i$ to monomer $i+1$ has rotational stiffness $K_i$. 
Recalling $\eTransformVec \coloneqq \trans{\nvecToDipole} \eFieldVec$, the Hamiltonian reads:
\begin{equation}\label{eq:wrm_hmltnn}
    \mathcal{H}= \sum_{i=1}^{n} \left(\frac{K_i}{2b}(\bm{\hat{n}}_{i+1}-\bm{\hat{n}}_{i})^2- \pmb{\mathcal{E}} \cdot \bm{\hat{n}}_i \right).
\end{equation}
where it is assumed that $K_n=0$ to simulate a hinged boundary condition (at the chain's end). 
In the limit of $b \rightarrow 0$ and $n \rightarrow \infty$, \eqref{eq:wrm_hmltnn} becomes:
\begin{equation}\label{wrm_hmltnn2}
     \mathcal{H}\simeq \int_{0}^L \left(\frac{K(s)}{2}\left|\frac{\df{\bm{{\hat{n}}}}}{\df{s}}\right|^2-\frac{1}{b} \pmb{\mathcal{E}} \cdot \bm{{\hat{n}}} \right) \df{s},
\end{equation}
where $L=nb$ is the chain length and the chain is parameterized by its arclength, $s \in \left[0,L\right]$. 
For simplicity we assume that the stiffness is same throughout the chain, i.e. $K(s)=K$.\\

Here we again consider a fixed end-to-end vector ensemble.
Formally, the end-to-end vector and normalization constraints are:
\begin{equation}\label{eq:WC_constraints}
\begin{split}
\bm{g}\left[\bm{\hat{n}}\right]&=\frac{1}{L}\left( b \sum_{i=1}^{n} \bm{\hat{n}}_{i}-\bm{r} \right)\simeq \frac{1}{L} \int_s \bm{{\hat{n}}}\ \df{s} -\frac{\bm{r}}{L}=\bm{0}, \\
h(s) &= \bm{\hat{n}}_i \cdot \bm{\hat{n}}_i - 1 \simeq \bm{\hat{n}}(s) \cdot \bm{\hat{n}}(s) - 1=0. 
\end{split}
\end{equation}
Let $\bm{g}\left[\bm{\hat{n}}\right]=(g_1,g_2,g_3)$, when evaluating the partition function, the constraints are enforced through Dirac delta distributions~\cite{lighthill1958introduction} and functionals~\cite{fredrickson2006equilibrium}: 
\begin{align*}
\partitionFunction=\int \exp[{-\beta  \mathcal{H}}]\ \delta({g_1}) \delta({g_2}) \delta({g_3}) \delta\left[{h}\right]\ \mathcal{D} \bm{\hat{n}}.
\end{align*}
%
Using the Fourier space representations of the Dirac delta distributions and functional~\cite{fredrickson2006equilibrium}, we have:
\begin{align}  \label{eq:prttn_eff}
\partitionFunction&=\frac{1}{(2\pi)^4} \int \exp\left[{-\beta  \mathcal{H}^{\eff}}\right]\  { \df^3 \bm{k}} \ \mathcal{D} \gamma \ \mathcal{D} \bm{{\hat{n}}},
\end{align}
 where $\bm{k}=\{k_1,k_2,k_3\}$ and $\gamma = \gamma\left(s\right)$ are wave vectors; and where the effective Hamiltonian is defined as:
 \begin{eqnarray} \label{eq:ham_eff}
 \begin{aligned}
    \mathcal{H}^{\eff}\simeq&\int_{0}^L \left(\frac{K}{2}\left|\frac{\df{\bm{{\hat{n}}}}}{\df{s}}\right|^2-\frac{1}{b} \pmb{\mathcal{E}} \cdot \bm{{\hat{n}}}  -\frac{i}{L \beta}\bm{k} \cdot \bm{{\hat{n}}}\right)\df{s}\\ &+\frac{i}{L \beta} \bm{k}\cdot \bm{r} -\int_{0}^L  \frac{i}{ L \beta}{\gamma}(s) \left(\bm{\hat{n}}\cdot \bm{\hat{n}} - 1\right)\ \df{s}.
 \end{aligned}
 \end{eqnarray}
The last term in \eqref{eq:ham_eff} is anharmonic, and therefore makes \eqref{eq:prttn_eff} difficult, if not impossible, to evaluate exactly.
As a result, we proceed to approximate the partition function by dropping the normalization constraint, i.e. we let $\gamma(s) \rightarrow 0$.
This elimination is equivalent to allowing the monomers to stretch or contract.  We show in \S 2 of the SI, by using a variational approach, that this assumption (interestingly) does not alter our result. \\


Next to evaluate (the harmonic approximation of) $\partitionFunction$, we use the Fourier transform of $\bm{\hat{n}}(s)$ given by:
\begin{equation} \label{eq:frr-1}
    \bm{\hat{n}}(s)=\sum_{q\in \mathbb{K}} \bm{\omega}(q)\exp{[i q s]},
\end{equation}
where $\mathbb{K}=\{q: q=2 n \pi / L, n\in \mathbb{Z} , q_{min}\leq |q| \leq q_{max} \}$ and $q_{min}=0$ and $q_{max}=2\pi / b$. 
The amplitudes of each mode $\bm{\omega}(q)$ can be achieved by:
\begin{equation}\label{eq:frr-2}
    \bm{\omega}(q)=\frac{1}{L}\int_{0}^L \bm{\hat{n}}(s) \exp{[-i q s]} \mathrm{d} s.
\end{equation}
By definitions \eqref{eq:frr-1} and \eqref{eq:frr-2}, we rewrite the effective Hamiltonian in \eqref{eq:ham_eff} as follows:
\begin{eqnarray} \label{eq:ham_eff_frr}
 \begin{aligned}
  \mathcal{H}^{\eff}_0=&  L \sum_{q \in \mathbb{K}} \left(\frac{K}{2} q^2 \left|\bm{\omega}(q)\right|^2\right)
  - \frac{L}{b} \pmb{\mathcal{E}} \cdot \bm{\omega}(0)\\
  &-\frac{i}{\beta} \bm{k} \cdot \bm{\omega}(0)
     +  \frac{i}{L \beta} \bm{k} \cdot \bm{r},
 \end{aligned}
 \end{eqnarray}
where $\mathcal{H}^{\eff}_0$ is the effective Hamiltonian upon dropping the normalization constraint, and $\pmb{\omega}(0)=\bm{r}/L$ by definition. 
Notice that the partition function is now harmonic in the unknowns.
Indeed, introducing $\bm{\alpha}({q,k})=[\omega_1(q_{min}:q_{max}),
\omega_2(q_{min}:q_{max}),
\omega_3(q_{min}:q_{max}),k_1,k_2,k_3]$ where $\Size\left(\bm{\alpha}\right)=:m $, we rewrite \eqref{eq:ham_eff_frr} as a quadratic function with the linear term:
\begin{eqnarray} \label{eq:ham_eff_def}
 \begin{aligned}
    \beta   \mathcal{H}^{\eff}_0=&\sum_{i,j=1}^{m} \frac{1}{2}\alpha_i {A}_{ij} \alpha_j +b_i  \alpha_i,
 \end{aligned}
 \end{eqnarray}
where $\bm{A}$ and $\bm{b}$ can be achieved from \eqref{eq:ham_eff_frr} by straight forward calculation. 
Plugging \eqref{eq:ham_eff_def} into \eqref{eq:prttn_eff}:
\begin{align}
\partitionFunction_0&=\frac{1}{(2\pi)^4}  \int \exp\left[{-\beta  \mathcal{H}^{\eff}_0}\right] 
 { \df^m \bm{\alpha}} ,\nonumber\\
&=\frac{1}{(2\pi)^4}\sqrt{  \frac{ (2\pi)^m}{ 
\prod_{|q| \in \mathbb{K} \& q\neq 0} (2L \beta {K} |q|^2)^3} }\exp\left[{\frac{\beta }{b} \pmb{\mathcal{E}} \cdot \bm{r}}\right].
\end{align}
 
 Once the partition function is obtained, we can identify the free energy of the system as follows:
 \begin{equation*}
    \cdGibbsChain=-\frac{1}{\beta}\ln\left({\partitionFunction_0}\right)=\text{const} + \frac{3}{2\beta}  \sum_{|q| \in \mathbb{K} \& q\neq 0} \ln(2 L \beta {K} |q|^2)-{\frac{ 1}{b} \pmb{\mathcal{E}} \cdot \bm{r}}
 \end{equation*}
 Thus recalling $\eTransformVec \coloneqq \trans{\nvecToDipole} \eFieldVec$, by definition, the polarization can be achieved by
 \begin{equation}\label{eq:polarization-free-energy}
 \bm{p}=-\frac{\partial \cdGibbsChain} { \partial \eFieldVec} = \bm{M}\frac{\bm{r}}{b}= n \chainStretch \nvecToDipole \rdir,
 \end{equation}
which is identical to \eqref{eq:plrztn-1st-apprch}.
We remark, as already noted earlier, that accounting for normalization constraint, does not change the polarization of the system.\\

In the remainder of the work, we show that the stretch-polarization relationship (i.e. the relationship between a chain's end-to-end vector and its net dipole) is a key component to the flexoelectric effect of elastomers consisting of polar monomers.
By \eqref{eq:plrztn-1st-apprch} and \eqref{eq:polarization-free-energy}, we can see that this relationship is consistent for both freely jointed and wormlike chains, and is invariant with respect to the ambient temperature or applied electric field.
Further, although the analytical expressions were determined by assuming negligible dipole-dipole interactions, Monte Carlo simulations agree nearly exactly, even when dipole-dipole interactions are significant (SI).
This surprisingly consistent result is a consequence of the monomer dipole constitutive response and the correspondence between thermodynamic state variables and ensemble averages.
Indeed, let $\avg{\generic}$ denote the ensemble average of $\generic$.
Then, the same relationship can be obtained almost by definition:
\begin{equation*}
\begin{split}
    \rvec &= \mLen \avg{\sum_{i=1}^{\n} \nvec_i}, \\
    \chainPolarVec = \avg{\sum_{i=1}^{\n} \dipoleVec_i} &= \nvecToDipole \avg{\sum_{i=1}^{\n} \nvec_i} = \nvecToDipole \frac{\rvec}{\mLen}.
\end{split}
\end{equation*}
The broad applicability of the monomer dipole constitutive model (i.e. dipolar molecules are not uncommon~\cite{lide2004crc}) and the consistency of the stretch-polarization relationship together suggest that the theory developed herein is potentially quite general; that is, it is expected to be valid for a wide range of types of elastomers, and for a variety of loading and environmental conditions.

\section{Network polarization: coarse-graining from chain-scale to the continuum scale}

In coarse-graining from the chain scale to the continuum scale, we assume that each material point can be represented by a probability distribution of chains and that these chains are in weak interaction with each other (that is, the free energy density at a material point is approximately the product of the average chain free energy--as if the chains were in isolation--and the density of chains per unit volume~\cite{tadmor2011modeling}). This assumption is standard in elastomer elasticity~\cite{treloar1975physics,boyce2000constitutive,james1943theory,flory1943jr,wu1993improved}, and has been used recently in elastomer electroelastictiy~\cite{cohen2016electroelasticity,cohen2016electromechanical,itskov2018electroelasticity,grasinger2020architected}. 
It is also consistent with our assumption that monomer-monomer interactions are negligible within a chain itself.
Let $\avgOverr{\generic}$ denote the average of $\generic$ over chains in the current configuration.
Then, by the weak interaction assumption, the closed-dielectric free energy density, $\cdGibbsDensity$, at a material point is
\begin{equation} \label{eq:cdGibbsDensity}
    \cdGibbsDensity = \chainDensity \avgOverr{\cdGibbsChain},
\end{equation}
where $\chainDensity$ is the number of chains per unit volume.
Similarly, we take the (continuum-scale) polarization (i.e. dipole moment per unit volume), $\polarVec$, as the average over chain polarizations; that is
\begin{equation} \label{eq:polar}
    \polarVec = \chainDensity \avgOverr{\chainPolarVec} = \chainDensity \nvecToDipole \avgOverr{\frac{\rvec}{\mLen}}.
\end{equation}
Note that, as a consequence of \eqref{eq:cdGibbsChain} and \eqref{eq:polar}, and since averaging is a linear operation: $\polarVec = -\partial \cdGibbsDensity / \partial \eFieldVec$~\footnote{This follows, in part, because their nonlocal relationship at the chain-scale simplifies to a local one~\cite{grasinger2020architected}.}.
This result, in congruence with a rigorous thermodynamic treatment of the consequences of deriving the chain scale free energy in a fixed electric field ensemble, leads us to the conclusion that $\cdGibbsDensity$ is the Legendre transform (with respect to polarization) of the Helmholtz free energy density, $\HelmholtzDensity$.
Thus,
\begin{equation} \label{eq:HelmholtzDensity}
    \HelmholtzDensity = \cdGibbsDensity - \polarVec \cdot \eFieldVec.
\end{equation}

The chain averaging considered thus far has been in the current configuration. To instead understand these quantities in terms of the reference configuration and continuum-scale deformations, we must make a kinematic assumption regarding how chains in the reference configuration are mapped into the current configuration\footnote{Historically, there has been some controversy related to this.
The two most common kinematic assumptions are what we will refer to as the \emph{affine deformation assumption} and the \emph{cooperative network assumption}.
In either case, the reference chains are arranged in a unit cell and the cell is deformed in relation to the deformation gradient, $\F$.
According to the affine deformation assumption, the chain end-to-end vectors in the reference configuration, $\Rvec$, are mapped to the current configuration by $\rvec = \F \Rvec$~\cite{treloar1975physics}; where as, for the cooperative network assumption, the unit cell is first rotated such that its axes are aligned with the principal frame~\cite{arruda1993threee}.
Then the cell is deformed along each of its axes according to its respective principal stretch.
In both the context of elasticity~\cite{treloar1975physics,boyce2000constitutive} and electroelasticity~\cite{grasinger2019multiscale,grasingerIPnongaussian}, there are important differences between the two assumptions and can sometimes even lead to qualitatively different behaviors.
In this work, we will focus on deformations which, in the frame of reference of interest, is such that $\F$ is diagonal.
Thus, the two kinematic assumptions will be equivalent.}.
Here we assume that chain end-to-end vectors are mapped from the reference to the current configuration under the deformation gradient, $\F$.
In the literature, this is known as the \emph{affine deformation assumption}~\cite{treloar1975physics}.\\

Finally, let $\avgOverR{\generic}$ denote the average of $\generic$ over chains in the reference configuration, $\J \coloneqq \det \F$, $\ChainDensity \coloneqq \J \chainDensity$ and $\PolarVec \coloneqq \J \polarVec$.
(We say that $\PolarVec$ is the pullback of $\polarVec$.)
Then
\begin{align*}
    \cdGibbsDensity\left(\F; \eFieldVec\right) &= \ChainDensity \avgOverR{\mathcal{F}\left(\F \Rvec, \eFieldVec\right)}, \\
    \HelmholtzDensity\left(\F; \eFieldVec\right) &= \cdGibbsDensity\left(\F, \eFieldVec\right) - \J^{-1} \PolarVec\left(\F\right) \cdot \eFieldVec, \\
    \PolarVec\left(\F\right) &= \ChainDensity \avgOverR{\chainPolarVec\left(\F \Rvec\right)},
\end{align*}
where we write $\PolarVec = \PolarVec\left(\F\right)$ to reinforce the fact that, in the case when monomer-monomer and chain-chain interactions are negligible, the polarization is determined uniquely by the deformation.
We also mention that $\eFieldVec$ is not a freely varying parameter but is determined from Gauss's law; that is,
\begin{equation*}
    \diverge{\eFieldVec} = \frac{1}{\vacPerm}\left(\chargeFree - \diverge{\polarVec}\right),
\end{equation*}
where $\vacPerm$ is the permittivity of free space and $\chargeFree$ is the density of free charges.\\

Before moving on, we pause to consider $\PolarVec$.
It is certainly true for an isotropic distribution of chains, but--since polymer chains are essentially entropic springs with a minimum energy length of $\rmag = 0$--even for a general stress-free distribution of chains, we expect that $\avgOverR{\Rvec} = 0$.
For any such reference distribution of chains, we find that, under the affine deformation assumption, the polarization vanishes for all $\F$.
Indeed, since averaging is linear,
\begin{equation*}
    \PolarVec = \ChainDensity \avgOverR{\nvecToDipole \frac{\F \Rvec}{\mLen}} = \frac{\ChainDensity}{\mLen} \nvecToDipole \F \avgOverR{\Rvec} = \bold{0}.
\end{equation*}
It would seem then that our model, at least thus far, is not very interesting. More precisely, this model does not give rise to \emph{piezoelectricity}. This is also reassuring in a sense that a conventional elastomer is indeed not piezoelectric, however, we will (in due course) aside from flexoelectricity, be able to comment on the precise conditions for the design of emergent piezoelectricity as well. 

\section{Flexoelectricity and the elastomer unit cell}\label{sec-flex}

\begin{figure*}[hbt!]
\begin{center}
\includegraphics[width=0.8\linewidth]{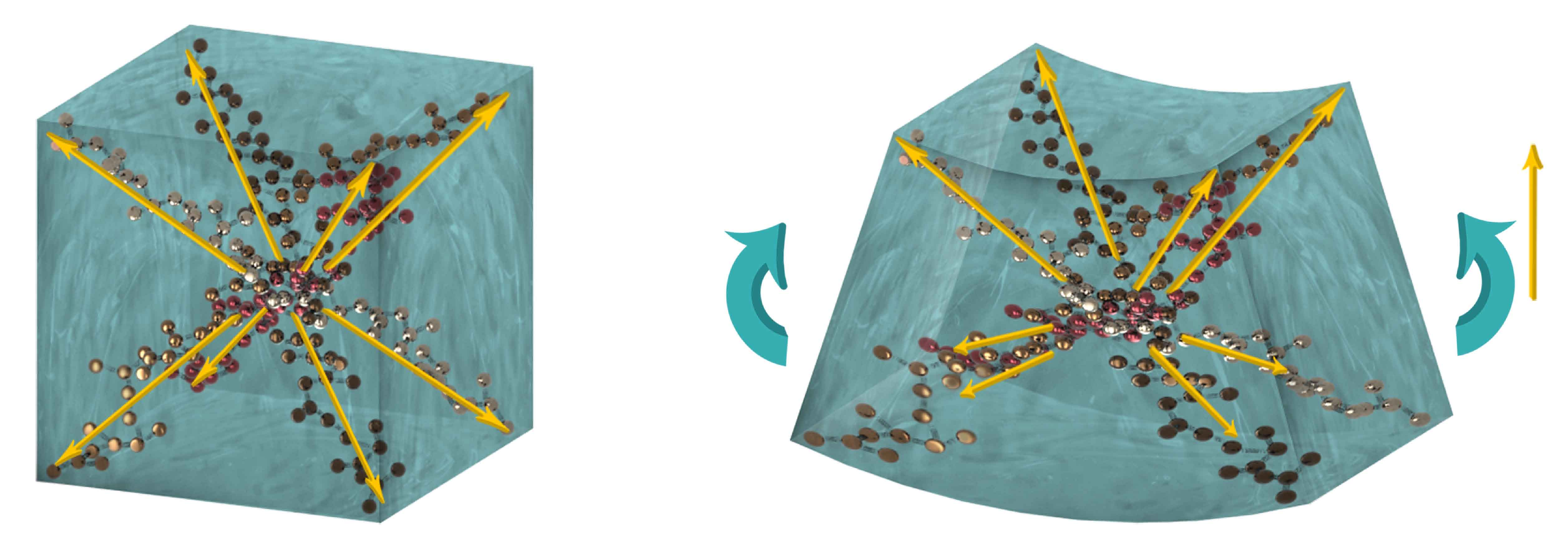}
\caption{\label{fig:ref-fig2} Schematic of a unit cell of an 8-chain model in (a) reference and (b) deformed state upon bending. The yellow arrows indicate the chain polarization in both configurations. While the unit cell in reference configuration is electrically neutral, the net polarization is upward in the deformed configuration under bending.}
\end{center}
\end{figure*}

Moving forward, we will be interested in the polarization as a function of $\takeGrad{\F}$ where $\F$ is the deformation gradient; that is, we are interested in $\PolarVec = \PolarVec\left(\takeGrad{\F}\right)$ and whether or not this relationship is as trivial as the polarization as a function of merely $\F$.
The $\takeGrad{\F}$ term could, for instance, be the result of a bending deformation and, hence, result in flexoelectricity.
To model the effects of $\takeGrad{\F}$ at a material point, we must now think beyond the probability distribution of $\Rvec$ and consider also the geometry of the unit cell.\\

Prior work (mentioned in the previous section~\cite{treloar1975physics,arruda1993threee,james1943theory,flory1943jr,wu1993improved}) which developed and made use of a unit cell in the reference configuration in order to relate continuum-scale deformations to chain-scale deformations were focused primarily on the statistical distribution of chain end-to-end vectors and the choice of kinematic assumption.
To the authors' knowledge, the geometry of the unit cell was not ever considered significant in and of itself (other than to ensure that $\Rmag = \mLen \sqrt{\n}$ for all $\Rvec$).
We are therefore in the position of being interested in defining a unit cell geometry but of being more or less without precedent.
Therefore, as a first example, we will assign dimensions to the Arruda-Boyce 8-chain model (or rather, determine the dimensions and let them, for the first time, have physical significance) (see~\cite{arruda1993threee,boyce2000constitutive} for example).
The unit cell \Fref{fig:ref-fig2}(a) consists of eight chains, each emanating from the center of the cell to one of the eight corners of the cube.\\

Each chain is assumed to satisfy $\Rmag = \mLen \sqrt{\n}$ \footnote{
The length $\mLen \sqrt{\n}$ is motivated by random walk statistics. 
It is the most likely distance (from the starting point) for a random walk of $\n$ steps with step length $\mLen$.
Assuming all chains in the reference configuration have length $\Rmag = \mLen \sqrt{\n}$ is a standard assumption in the constitutive modeling of rubber~\cite{treloar1975physics,arruda1993threee, james1943theory,flory1943jr,wu1993improved}.
}; thus, the length of an edge of the unit cell is $\cellLength \coloneqq 2\mLen\sqrt{\n / 3}$.
The kinematic assumption that we now make--similar to the affine deformation assumption--will be that 1) the unit cell deforms under $\F$ (where $\F$ is allowed to vary throughout the cell) and 2) the beginning and end of the chain end-to-end vectors remain rigidly attached to their corresponding points in the unit cell.
For the 8-chain model, this means that each of the end-to-end vectors, in the deformed configuration, still begins at the center of the cell and ends at its respective corner.\\

\newcommand{\auxrz}{\alpha_3}
Having decided on an example unit cell, we next identify deformations of interest.
As mentioned previously, we will limit our investigation, for the time being, to diagonal $\F$.
In addition, since most elastomers are approximately incompressible, we will consider $\F$ such that $\J = 1$.
Thus, let 
\begin{equation} \label{def-grad}
\F\left(\xvec\right) = \diag\left(\pStretch{1}\left(\xvec\right), \pStretch{2}\left(\xvec\right), \left( \pStretch{1}\left(\xvec\right)\pStretch{2}\left(\xvec\right)\right)^{-1}\right),
\end{equation} where $\xvec$ is the position of a material point in the reference configuration.
For ease of notation, let $\dpStretch{i}{j} \coloneqq \partial \pStretch{i} / \partial \xj_{j}$.
Then, we assume $\F$ varies gradually enough relative to the size of a unit cell such that $\takeGrad{\F}$ is approximately constant.
To be precise, we assume $\smallpar \coloneqq \max_{i, j} \left|\Rmag \dpStretch{i}{j}\right| \ll 1$.
Next let $\yvec$ denote a position within the 8-chain unit cell so that $\yj_i \in \left[\xj_i - \cellLength / 2, \xj_i + \cellLength/ 2\right]$.
Then, $\pStretch{i}\left(\yvec\right) = \pStretch{i}\left(\xvec\right) + \sum_{j=1}^3 \dpStretch{i}{j}\left(\xvec\right) \yj_j + \orderOf{\smallpar^2}$.
Neglecting $\orderOf{\smallpar^2}$ terms, we have the result:
\begin{equation}
	\avgOverr{\rvec} = \frac{\Rmag}{3} \left(
	 \Rmag \dpStretch{1}{1},
	 \Rmag \dpStretch{2}{2},
	 \auxrz
	 \right),
\end{equation}
where
\begin{equation*}
\begin{split}
\auxrz &= \frac{\sqrt{3}}{8} \sum_{i=1}^2 \sum_{j=1}^2 \sum_{k=1}^2 \Bigg\{\left(-1\right)^{k} \\
&\times \left[\pStretch{1} + \frac{\Rmag}{\sqrt{3}} \left(\left(-1\right)^{i} \dpStretch{1}{1} + \left(-1\right)^{j} \dpStretch{1}{2} + \left(-1\right)^{k} \dpStretch{1}{3}\right)\right]^{-1} \\
&\times \left[\pStretch{2} + \frac{\Rmag}{\sqrt{3}} \left(\left(-1\right)^{i} \dpStretch{2}{1} + \left(-1\right)^{j} \dpStretch{2}{2} + \left(-1\right)^{k} \dpStretch{2}{3}\right)\right]^{-1} \Bigg\}
\end{split}
\end{equation*}
The expression for $\auxrz$ is still complicated however.
To simplify further, and because we are ultimately interested in flexoelectricity, consider $\pStretch{1}$ and $\pStretch{2}$ of the form:
\newcommand{\auxPZero}[1]{\pZero_{#1}}
\begin{equation}\label{eq:bend}
	\pStretch{j} = \auxPZero{j} + \dpStretch{j}{3} \xj_{3},
\end{equation}
as this could correspond to a bending about some axis in the $\euclid{1}, \euclid{2}$ plane. 
In this case,
\begin{equation} \label{eq:bending-avgrvec}
	\avgOverr{\rvec} = \left(0, 0, \frac{-3 \Rmag^2\left(\auxPZero{2}\dpStretch{1}{3} + \auxPZero{1}\dpStretch{2}{3}\right)}{\left(3 {\auxPZero{1}}^2 - \Rmag^2 \dpStretch{1}{3}^2\right)\left(3 {\auxPZero{2}}^2 - \Rmag^2 \dpStretch{2}{3}^2\right)}\right),
\end{equation}
such that, in general, we have a nonzero polarization.
Further, we can see from \eqref{eq:polar} that if $\nvecToDipole = \dipoleMag \identity$ (where $\identity$ is the identity operator) then $\polarVec$ is aligned (or anti-aligned) with $\takeGrad{\pStretch{1}}$ and $\takeGrad{\pStretch{2}}$ (i.e. the $\euclid{3}$ direction).\\

We can make the following observations regarding \eqref{eq:bending-avgrvec}:
\begin{itemize}
	\item It is interesting to see in \eqref{eq:bending-avgrvec} that $\avgOverr{\rvec}$ can diverge if $3 {\auxPZero{1}}^2 = \Rmag^2 \dpStretch{1}{3}^2$ or $3 {\auxPZero{2}}^2 = \Rmag^2 \dpStretch{2}{3}^2$.
	For this to occur, roughly speaking, the deformation gradient $\F$ must change appreciably relative to the size of the unit cell (more precisely, $\Rmag \dpStretch{i}{3} = \orderOf{1}$).
	This violates a prior assumption and so it should not be surprising that it leads to nonphysical results.
	If $\F$ changes appreciably relative to size of the unit cell then perhaps our unit cell is too large or we need to resort to a different modeling approach altogether.
	\item It is also interesting that if $\auxPZero{2}\dpStretch{1}{3} = - \auxPZero{1}\dpStretch{2}{3}$ then $\PolarVec$ vanishes.
	This is, in part, because of incompressibility.
	Indeed, let $\pStretch{3}$ denote the stretch in the $\euclid{3}$ direction.
	Then,
	\begin{equation*}
	    \pStretch{3} = \left(\auxPZero{1}\auxPZero{2} + \left(\auxPZero{2}\dpStretch{1}{3} + \auxPZero{1}\dpStretch{2}{3}\right) \xj_{3} + \dpStretch{1}{3} \dpStretch{2}{3} \xj_3^2\right)^{-1},
	\end{equation*}
	such that $\left(\auxPZero{2}\dpStretch{1}{3} + \auxPZero{1}\dpStretch{2}{3}\right) = 0$ implies $\pStretch{3}\left(\xj_3\right)$ is an even function; and therefore all of the chains experience the same stretch in the $\euclid{3}$ direction.
	\item To see how the flexoelectric effect scales with stretch, consider $\auxPZero{3} \coloneqq \left(\auxPZero{1}\auxPZero{2}\right)^{-1}$.
	Neglecting $\orderOf{\smallpar^2}$ terms in the denominator of \eqref{eq:bending-avgrvec} results in $\PolarMag_3 \sim \orderOf{\left(\auxPZero{2}\dpStretch{1}{3} + \auxPZero{1}\dpStretch{2}{3}\right) \auxPZero{3}^{2}}$.
	As a result, we see that our theory agrees with the surprising experimental result that pre-stretching the elastomer in the direction of $\takeGrad{\F}$ can lead to giant flexoelectricity~\cite{baskaran2011experimental}.
	Note that if $\auxPZero{1} = \auxPZero{2} = \auxPZero{3}^{-1/2}$ then this enhancement scales as $\auxPZero{3}^{3/2}$; however, if $1 = \auxPZero{1} \gg \auxPZero{2} = \auxPZero{3}^{-1}$ (or vice versa) then \emph{the enhancement is quadratic in $\auxPZero{3}$}.
\end{itemize}

\section{Elastomer design and the conditions for the \emph{piezoelectric} effect}\label{sec-piezo}

\begin{figure}
    \centering
    \includegraphics[width=0.7\linewidth]{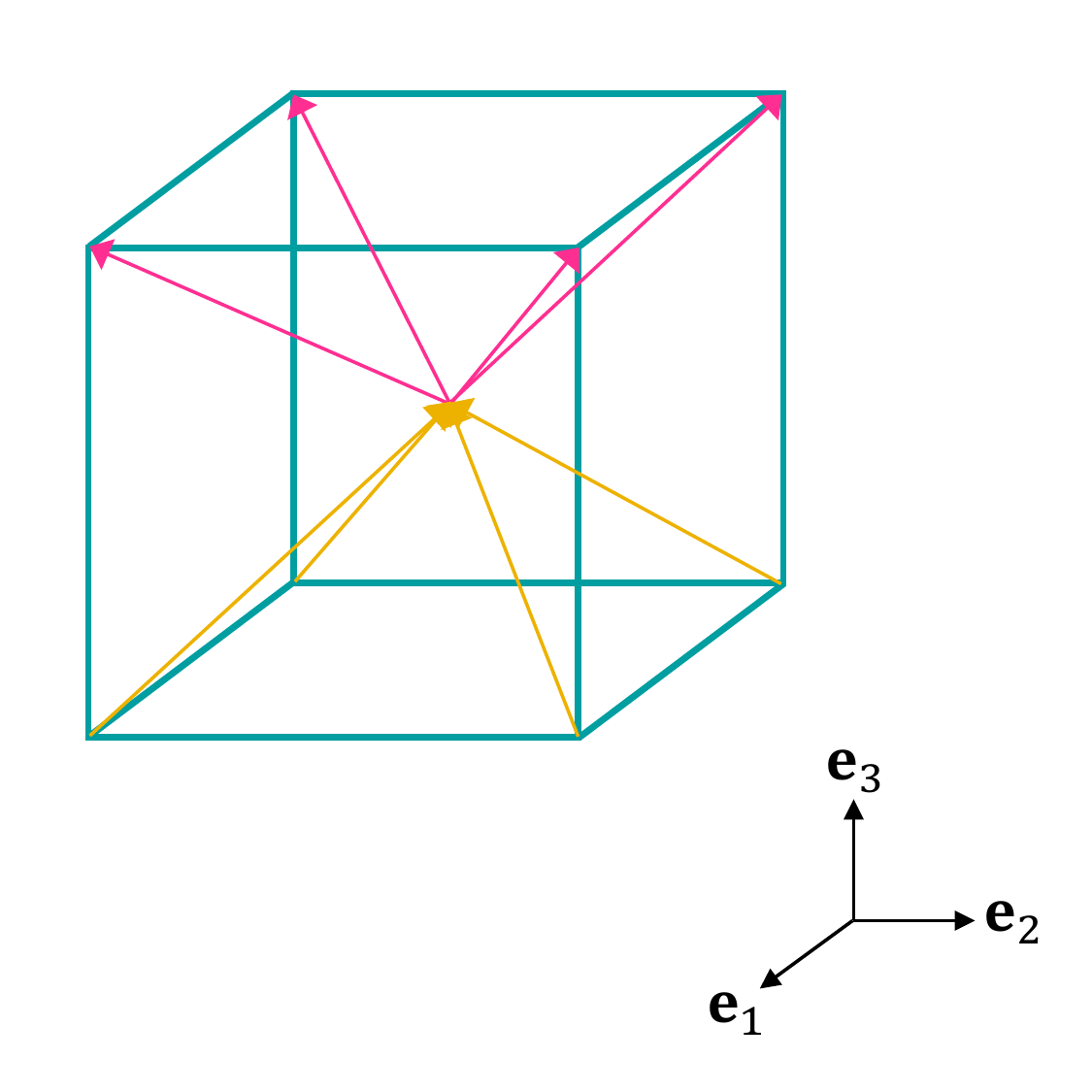}
    \caption{\small Schematic of an eight-chain model exhibiting piezoelectric behavior. The arrows show the direction of the polarization of each chain.
    }
    \label{fig:piezo}
\end{figure}

Although not the primary focus of the current work, since piezoelectricity is arguably the most well-known and direct form of electromechanical coupling, it would be illuminating to understand the precise conditions under which an elastomer can be designed to exhibit such a property. We emphasize once again that while flexoelectricity is universal, piezoelectricity is not and we know of no piezoelectric elastomers\footnote{Certain polymers--but not elastomers--do exhibit piezoelectricity such as polyvinylidene difluoride(PVDF) \cite{fukada1969piezoelectric} and of polymer-ceramic
composites \cite{sappati2018piezoelectric}.}. To make progress on this, we assume that $\nvecToDipole$ differs from chain to chain. 
By doing so, from  \eqref{eq:polar} the polarization becomes:

 \begin{eqnarray}\label{eq:peizo}
    \PolarVec =  \J \polarVec = \J \chainDensity \avgOverr{\chainPolarVec} = \ChainDensity \avgOverr{ \nvecToDipole \frac{\rvec}{\mLen}}. 
\end{eqnarray}
 
A special case can be constructed by considering a relation of  $\nvecToDipole^{(2)}= -\nvecToDipole^{(1)} $ in which $\nvecToDipole^{(1)}$($\nvecToDipole^{(2)}$) belongs to the upper(lower) half chains in $\euclid{3}$ direction of the eight chain model as shown in Fig~\ref{fig:piezo}. \\


Here, by considering the eight chain model, we are assuming that the system is isotropic from a mechanical (elastic) perspective, while the linear transformation $\nvecToDipole$ is not unique in the entire cell and thus not isotropic from an electrostatic point of view.
This assumption gives us a nonvanishing polarization, even in the reference configuration. \\

We consider a deformation which varies along the unit cell i.e. $\pStretch{i}\left(\yvec\right) = \pStretch{i}\left(\xvec\right) + \sum_{j=1}^3 \dpStretch{i}{j}\left(\xvec\right) \yj_j + \orderOf{\smallpar^2}$. Based on the model we introduced for flexoelectricity in previous section, we obtain:
 \begin{equation*}
 \begin{split}
	\avgOverr{\bm{M}\rvec} &=\frac{1}{2}\left[ \avgOverr{\bm{M}^{(1)} \rvec} + \avgOverr{\bm{M}^{(2)}  \rvec} \right]\\
	&=
	\bm{M}^{(1)} \frac{\Rmag}{3} \left(
	 0,
	 0,
	 \auxrz
	 \right),
 \end{split}
\end{equation*}
where
\begin{equation*}
\begin{split}
\auxrz & = \frac{\sqrt{3}}{8} \sum_{i=1}^2 \sum_{j=1}^2 \sum_{k=1}^2 \\
&\Bigg\{
\left[\pStretch{1} + \frac{\Rmag}{\sqrt{3}} \left(\left(-1\right)^{i} \dpStretch{1}{1} + \left(-1\right)^{j} \dpStretch{1}{2} + \left(-1\right)^{k} \dpStretch{1}{3}\right)\right]^{-1} \\
&\times \left[\pStretch{2} + \frac{\Rmag}{\sqrt{3}} \left(\left(-1\right)^{i} \dpStretch{2}{1} + \left(-1\right)^{j} \dpStretch{2}{2} + \left(-1\right)^{k} \dpStretch{2}{3}\right)\right]^{-1} \Bigg\}.
\end{split}
\end{equation*}

Again, our ultimate goal is to investigate flexoelectricity. Therefore, restricting our attention to the bending about some axis in the $\euclid{1}, \euclid{2}$ plane same as \eqref{eq:bend}, we obtain
the net polarization after the deformation of the unit cell as follows:
 \begin{eqnarray*}
      \begin{aligned}
    \PolarVec &= \ChainDensity \avgOverr{ \nvecToDipole \frac{\rvec}{\mLen}}
    \\ &=\ChainDensity \nvecToDipole^{(1)} \left(0, 0, \frac{\sqrt{3} \Rmag\left(3\auxPZero{1}\auxPZero{2}+ \Rmag^2 \dpStretch{1}{3}\dpStretch{2}{3} \right)}{\left(3 {\auxPZero{1}}^2 - \Rmag^2 \dpStretch{1}{3}^2\right)\left(3 {\auxPZero{2}}^2 - \Rmag^2 \dpStretch{2}{3}^2\right)}\right)^T.\\
\end{aligned}
\end{eqnarray*}

Here, there are few observations that can be made: 
\begin{enumerate}
\item The net polarization after the deformation has contributions from two different types of deformations. The first term (in the numerator) is from the deformation gradient itself which is indication of the piezoelectric effect, meanwhile the second term originates from the gradients of the deformation gradient which is indication of the flexoelectric behavior. 
\item By assumption, $\max_j \left|\Rmag \dpStretch{j}{3}\right| = \orderOf{\epsilon}$ is small.
Thus, the second term, which is attributed to the flexoelectric effect, is much smaller than the flexoelectric response of the electrically isotropic case in previous section 
(to be precise, it is $\orderOf{\epsilon^2}$ compared to $\orderOf{\epsilon}$).
The conclusion is that, in this case, while we have increased the piezoelectric effect, it has been at the expense of flexoelectricity.
\item Similar to the previous setup of the unit cell, stretching the unit cell in $\euclid{3}$ direction, can enhance both piezoelectric and flexoelectric response in a linear and quadratic way respectively.
\end{enumerate}

\section{The design of elastomers for emergent flexoelectricity}

Clearly, we should maximize the spectral radius of $\nvecToDipole$ if we aim to maximize the flexoelectric effect.
For the case of a fixed dipole rigidly attached to each monomer (i.e. $\nvecToDipole = \dipoleMag \tens{Q}$), this amounts to maximizing the magnitude of the dipoles, $\dipoleMag$.
We could potentially engage in such a design process (computationally) by using density functional theory (see ~\cite{ashraf2009theoretical}, for example).
However, we also note that it is often trivial to increase the density of cross-links in an elastomer.
In addition, we could theoretically influence the alignment of chains in the network by weakly cross-linking, applying an external electric field, and then cross-linking further.
Therefore, it is also conceivable that the unit cell itself be taken as a design variable (see ~\cite{grasinger2020architected}).
This may prove to be a fruitful endeavor since, for elastomers that consist of the type of polymer chains considered herein, the geometry of the unit cell and the arrangement of chains within the unit cell influence the magnitude and direction of the flexoelectric effect.\\

\begin{figure}
    \centering
    \includegraphics[width=0.7\linewidth]{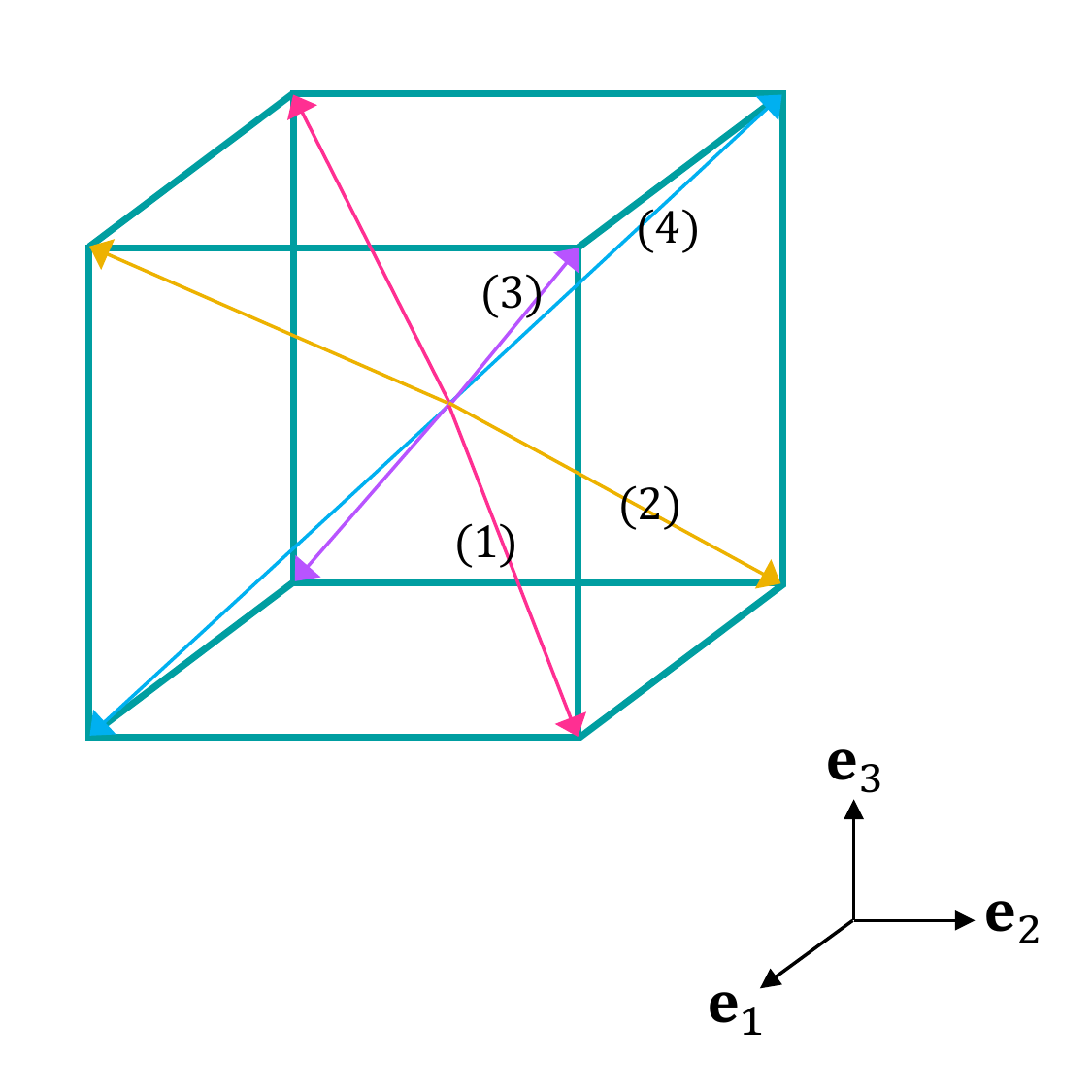}
    \caption{\small Design of an eight-chain unit cell to tune the flexoelectric response.
    Polymers are paired together through an inversion symmetry as shown by numbers; and, when the electrical properties of the four pairs are tuned to maximize the gradient of the quadruple moment of the unit cell, the flexoelectric response is stretch-invariant.
    }
    \label{fig:quadru}
\end{figure}

{
The main goal that we now focus on is to design an elastomer unit cell to tune the flexoelectric response. In particular, we propose that (in analogy with the dominant role a dipole moment plays in the phenomenon of piezoelectricity) that flexoelectricity is primarily dictated by the next-order multipole expansion term--the quadrupolar moment. In particular, we anticipate unit cells with large variations to have a stronger flexoelectric response. The quadrupole moment can be constructed from two dipole moments in opposite direction
. The quadrupole moment($\mathcal{Q}$) is a second order moment of charges and can be formulated as \cite{griffiths2005introduction}

\begin{equation}
\mathcal{Q}=\sum_{i=1}^{n} \left [3 \bold{x}_i \otimes \bold{x}_i - |\bold{x}_i|^2 \mathbb{I}  \right ] q_i,
\end{equation}
where the sum is over all discrete charges($q_i$), $\bold{x}_i$ is the position vector of the charge $q_i$ and $\mathbb{I}$ is the identity matrix.
Analogous to the fact that dipole distribution can be architected to design materials with high piezoelectricity, quadrupole moment and its gradient can play the same role in designing giant flexoelectric materials. \\

Here we aim to establish a reference unit cell with vanishing net polarization but finite quadrupole moment and its gradient. The authors claim that the higher gradient in quadrupole moment will result in higher flexoelectric effect. The simplest example can be achieved by enforcing an inversion symmetry about the center of the 8-chain unit cell,
(as shown in \Fref{fig:quadru})
which leaves four pairs of chains each with a different $\bm{M}$ and each acting as a quadrupole moment.}

We make the choice:
\begin{equation*}
\bm{M}=\frac{1}{K_1} \bm{M}^{(1)}= \frac{1}{K_2}  \bm{M}^{(2)}=\frac{1}{K_3}  \bm{M}^{(3)}=\frac{1}{K_4}  \bm{M}^{(4)},
\end{equation*}
where, $K_i$($i=1:4$) are some arbitrary real numbers. 
Again consider a deformation gradient of the form \eqref{def-grad} and its associated assumptions.
By restricting the deformation to the bending of the unit cell similar to \eqref{eq:bend}, the polarization after the deformation is obtained as:
 \begin{eqnarray*}\label{eq:quad-idea}
 \begin{aligned}
    \PolarVec &
    = \sum_i \frac{\ChainDensity^{(i)}}{\mLen} \avgOverr{ \nvecToDipole^{(i)}  \rvec^{(i)}}
    =\frac{\ChainDensity}{b}\bm{M}^{(1)} {\Rmag^2} \left(
	  \alpha_1,
	   \alpha_2,
	  \alpha_3
	 \right),
	 \end{aligned}
\end{eqnarray*}
where $\alpha_i$ are defined as:
\begin{eqnarray*}
\begin{split}
\alpha_1&=\frac{1}{12}(K_2+K_3-K_1-K_4)\lambda_{1,3},\\
\alpha_2&=\frac{1}{12}(K_3+K_4-K_1-K_2)\lambda_{2,3},\\
\alpha_3&= \frac{-3 (K_1+K_2+K_3+K_4) \left(\auxPZero{2}\dpStretch{1}{3} + \auxPZero{1}\dpStretch{2}{3}\right)}{4 \left(3 {\auxPZero{1}}^2 - \Rmag^2 \dpStretch{1}{3}^2\right)\left(3 {\auxPZero{2}}^2 - \Rmag^2 \dpStretch{2}{3}^2\right)}.
\end{split}
\end{eqnarray*}
Upon a closer examination, we notice that the third term is exactly the flexoelectricity constant we obtained in \eqref{eq:bending-avgrvec} for the case $K_1=K_2=K_3=K_4=1$, which is equivalent to the electrically isotropic case.\\

However, if we allow the $K_i$ to vary, we see that we can obtain flexoelectricity in the plane orthogonal to $\euclid{3}$.
Indeed, if we make the choice is $K_1=K_4=-K_2=-K_3$, this results in enhancing the $\alpha_1$ while eliminating $\alpha_3$.
Interestingly, we see that, in this case, the flexoelectric effect is invariant with respect to $\auxPZero{1}$ and $\auxPZero{2}$.
This property may be particularly useful for applications in soft sensors and robotics where we may expect finite stretches to occur but would prefer the flexoelectric response to remain constant, or, at the very least, not diminish.

\section{Concluding remarks}

In this work, we have established the mechanisms (and the underlying theory) underpinning flexoelectricity in elastomers. Surprisingly, our theory shows that giant flexoelectricity can be attained in incompressible elastomers by pre-stretching the material in the strain gradient direction. In particular, the scaling of the flexoelectric effect with respect to pre-stretch is at least super-linear (with exponent $3/2$) and, for certain loading conditions, can scale quadratically with pre-stretch. This suggests a facile route to achieve high-output energy harvesting and high-fidelity soft sensors.\\

We also considered the design of the polymer network architecture. By poling the chains (e.g. using an applied electric field), we can develop a piezoelectric elastomer. This piezoelectricity is at expense of the flexoelectric effect which diminishes in the poled architecture.
The emergent flexoelectric effect can also be (1) made stretch-invariant, and (2) tuned in different directions relative to the direction of the strain gradient by adjusting the quadrupolar moment of the material unit cell. Stretch-invariance may prove useful for soft sensors, and the tunable nature of the flexoelectric direction may be useful for producing soft robots with larger degree of freedom actuators.\\

\acknow{Authors K.M. and P.S. were supported by the University of Houston and the M.D. Anderson Professorship. P.S. and M.G. gratefully acknowledge encouragement and discussions by Professor Kaushik Dayal. M.G. would also like to thank Professors Prashant Purohit and Gal de Botton for insightful discussions.}

\showacknow 

\pnasbreak


\begin{thebibliography}{98}
\providecommand{\natexlab}[1]{#1}
\providecommand{\url}[1]{\texttt{#1}}
\expandafter\ifx\csname urlstyle\endcsname\relax
  \providecommand{\doi}[1]{doi: #1}\else
  \providecommand{\doi}{doi: \begingroup \urlstyle{rm}\Url}\fi

\bibitem[Keplinger et~al.(2012)Keplinger, Li, Baumgartner, Suo, and
  Bauer]{keplinger2012harnessing}
Christoph Keplinger, Tiefeng Li, Richard Baumgartner, Zhigang Suo, and
  Siegfried Bauer.
\newblock Harnessing snap-through instability in soft dielectrics to achieve
  giant voltage-triggered deformation.
\newblock \emph{Soft Matter}, 8\penalty0 (2):\penalty0 285--288, 2012.

\bibitem[Sun et~al.(2012)Sun, Zhao, Illeperuma, Chaudhuri, Oh, Mooney, Vlassak,
  and Suo]{sun2012highly}
Jeong-Yun Sun, Xuanhe Zhao, Widusha~RK Illeperuma, Ovijit Chaudhuri, Kyu~Hwan
  Oh, David~J Mooney, Joost~J Vlassak, and Zhigang Suo.
\newblock Highly stretchable and tough hydrogels.
\newblock \emph{Nature}, 489\penalty0 (7414):\penalty0 133--136, 2012.

\bibitem[Kwak et~al.(2020)Kwak, Han, Xie, Chung, Lee, Avila, Yohay, Chen,
  Liang, Patel, et~al.]{kwak2020wireless}
Jean~Won Kwak, Mengdi Han, Zhaoqian Xie, Ha~Uk Chung, Jong~Yoon Lee, Raudel
  Avila, Jessica Yohay, Xuexian Chen, Cunman Liang, Manish Patel, et~al.
\newblock Wireless sensors for continuous, multimodal measurements at the skin
  interface with lower limb prostheses.
\newblock \emph{Science translational medicine}, 12\penalty0 (574), 2020.

\bibitem[Rogers et~al.(2010)Rogers, Someya, and Huang]{rogers2010materials}
John~A Rogers, Takao Someya, and Yonggang Huang.
\newblock Materials and mechanics for stretchable electronics.
\newblock \emph{science}, 327\penalty0 (5973):\penalty0 1603--1607, 2010.

\bibitem[Zhang et~al.(2020)Zhang, Liang, and Rogers]{zhang2020water}
Qian Zhang, Qijie Liang, and John~A Rogers.
\newblock Water-soluble energy harvester as a promising power solution for
  temporary electronic implants.
\newblock \emph{APL Materials}, 8\penalty0 (12):\penalty0 120701, 2020.

\bibitem[Zhao et~al.(2019)Zhao, Kim, Chester, Sharma, and
  Zhao]{zhao2019mechanics}
Ruike Zhao, Yoonho Kim, Shawn~A Chester, Pradeep Sharma, and Xuanhe Zhao.
\newblock Mechanics of hard-magnetic soft materials.
\newblock \emph{Journal of the Mechanics and Physics of Solids}, 124:\penalty0
  244--263, 2019.

\bibitem[Ze et~al.(2020)Ze, Kuang, Wu, Wong, Montgomery, Zhang, Kovitz, Yang,
  Qi, and Zhao]{ze2020magnetic}
Qiji Ze, Xiao Kuang, Shuai Wu, Janet Wong, S~Macrae Montgomery, Rundong Zhang,
  Joshua~M Kovitz, Fengyuan Yang, H~Jerry Qi, and Ruike Zhao.
\newblock Magnetic shape memory polymers with integrated multifunctional shape
  manipulation.
\newblock \emph{Advanced Materials}, 32\penalty0 (4):\penalty0 1906657, 2020.

\bibitem[Xu et~al.(2014)Xu, Zhang, Jia, Mathewson, Jang, Kim, Fu, Huang, Chava,
  Wang, et~al.]{xu2014soft}
Sheng Xu, Yihui Zhang, Lin Jia, Kyle~E Mathewson, Kyung-In Jang, Jeonghyun Kim,
  Haoran Fu, Xian Huang, Pranav Chava, Renhan Wang, et~al.
\newblock Soft microfluidic assemblies of sensors, circuits, and radios for the
  skin.
\newblock \emph{Science}, 344\penalty0 (6179):\penalty0 70--74, 2014.

\bibitem[Kang et~al.(2016)Kang, Murphy, Hwang, Lee, Harburg, Krueger, Shin,
  Gamble, Cheng, Yu, et~al.]{kang2016bioresorbable}
Seung-Kyun Kang, Rory~KJ Murphy, Suk-Won Hwang, Seung~Min Lee, Daniel~V
  Harburg, Neil~A Krueger, Jiho Shin, Paul Gamble, Huanyu Cheng, Sooyoun Yu,
  et~al.
\newblock Bioresorbable silicon electronic sensors for the brain.
\newblock \emph{Nature}, 530\penalty0 (7588):\penalty0 71--76, 2016.

\bibitem[Kim et~al.(2012)Kim, Wang, Keum, Ghaffari, Kim, Tao, Panilaitis, Li,
  Kang, Omenetto, et~al.]{kim2012thin}
Dae-Hyeong Kim, Shuodao Wang, Hohyun Keum, Roozbeh Ghaffari, Yun-Soung Kim,
  Hu~Tao, Bruce Panilaitis, Ming Li, Zhan Kang, Fiorenzo Omenetto, et~al.
\newblock Thin, flexible sensors and actuators as ‘instrumented’surgical
  sutures for targeted wound monitoring and therapy.
\newblock \emph{Small}, 8\penalty0 (21):\penalty0 3263--3268, 2012.

\bibitem[Kim et~al.(2011)Kim, Lu, Ma, Kim, Kim, Wang, Wu, Won, Tao, Islam,
  et~al.]{kim2011epidermal}
Dae-Hyeong Kim, Nanshu Lu, Rui Ma, Yun-Soung Kim, Rak-Hwan Kim, Shuodao Wang,
  Jian Wu, Sang~Min Won, Hu~Tao, Ahmad Islam, et~al.
\newblock Epidermal electronics.
\newblock \emph{science}, 333\penalty0 (6044):\penalty0 838--843, 2011.

\bibitem[Han et~al.(2019)Han, Wang, Yang, Liang, Bai, Yan, Li, Xue, Wang, Akar,
  et~al.]{han2019three}
Mengdi Han, Heling Wang, Yiyuan Yang, Cunman Liang, Wubin Bai, Zheng Yan, Haibo
  Li, Yeguang Xue, Xinlong Wang, Banu Akar, et~al.
\newblock Three-dimensional piezoelectric polymer microsystems for vibrational
  energy harvesting, robotic interfaces and biomedical implants.
\newblock \emph{Nature Electronics}, 2\penalty0 (1):\penalty0 26--35, 2019.

\bibitem[Erturk and Inman(2011)]{erturk2011piezoelectric}
Alper Erturk and Daniel~J Inman.
\newblock \emph{Piezoelectric energy harvesting}.
\newblock John Wiley \& Sons, 2011.

\bibitem[Nan et~al.(2018)Nan, Kang, Li, Yu, Zhu, Wang, Dunn, Zhou, Xie, Agne,
  et~al.]{nan2018compliant}
Kewang Nan, Stephen~Dongmin Kang, Kan Li, Ki~Jun Yu, Feng Zhu, Juntong Wang,
  Alison~C Dunn, Chaoqun Zhou, Zhaoqian Xie, Matthias~T Agne, et~al.
\newblock Compliant and stretchable thermoelectric coils for energy harvesting
  in miniature flexible devices.
\newblock \emph{Science advances}, 4\penalty0 (11):\penalty0 eaau5849, 2018.

\bibitem[Kim et~al.(2009)Kim, Zhao, Jang, Lee, Kim, Kim, Ahn, Kim, Choi, and
  Hong]{kim2009large}
Keun~Soo Kim, Yue Zhao, Houk Jang, Sang~Yoon Lee, Jong~Min Kim, Kwang~S Kim,
  Jong-Hyun Ahn, Philip Kim, Jae-Young Choi, and Byung~Hee Hong.
\newblock Large-scale pattern growth of graphene films for stretchable
  transparent electrodes.
\newblock \emph{nature}, 457\penalty0 (7230):\penalty0 706--710, 2009.

\bibitem[Khang et~al.(2006)Khang, Jiang, Huang, and
  Rogers]{khang2006stretchable}
Dahl-Young Khang, Hanqing Jiang, Young Huang, and John~A Rogers.
\newblock A stretchable form of single-crystal silicon for high-performance
  electronics on rubber substrates.
\newblock \emph{Science}, 311\penalty0 (5758):\penalty0 208--212, 2006.

\bibitem[Xu et~al.(2013)Xu, Zhang, Cho, Lee, Huang, Jia, Fan, Su, Su, Zhang,
  et~al.]{xu2013stretchable}
Sheng Xu, Yihui Zhang, Jiung Cho, Juhwan Lee, Xian Huang, Lin Jia, Jonathan~A
  Fan, Yewang Su, Jessica Su, Huigang Zhang, et~al.
\newblock Stretchable batteries with self-similar serpentine interconnects and
  integrated wireless recharging systems.
\newblock \emph{Nature communications}, 4\penalty0 (1):\penalty0 1--8, 2013.

\bibitem[Kim et~al.(2018)Kim, Yuk, Zhao, Chester, and Zhao]{kim2018printing}
Yoonho Kim, Hyunwoo Yuk, Ruike Zhao, Shawn~A Chester, and Xuanhe Zhao.
\newblock Printing ferromagnetic domains for untethered fast-transforming soft
  materials.
\newblock \emph{Nature}, 558\penalty0 (7709):\penalty0 274--279, 2018.

\bibitem[Kim and Tadokoro(2007)]{kim2007electroactive}
Kwang~J Kim and Satoshi Tadokoro.
\newblock Electroactive polymers for robotic applications.
\newblock \emph{Artificial Muscles and Sensors}, 23:\penalty0 291, 2007.

\bibitem[Kim et~al.(2019)Kim, Parada, Liu, and Zhao]{kim2019ferromagnetic}
Yoonho Kim, German~A Parada, Shengduo Liu, and Xuanhe Zhao.
\newblock Ferromagnetic soft continuum robots.
\newblock \emph{Science Robotics}, 4\penalty0 (33), 2019.

\bibitem[Sessler and Hillenbrand(1999)]{sessler1999electromechanical}
GM~Sessler and J~Hillenbrand.
\newblock Electromechanical response of cellular electret films.
\newblock In \emph{10th International Symposium on Electrets (ISE 10).
  Proceedings (Cat. No. 99 CH36256)}, pages 261--264. IEEE, 1999.

\bibitem[Sessler(1980)]{sessler1980physical}
Gerhard~Martin Sessler.
\newblock Physical principles of electrets.
\newblock In \emph{Electrets}, pages 13--80. Springer, 1980.

\bibitem[Wegener and Bauer(2005)]{wegener2005microstorms}
Michael Wegener and Siegfried Bauer.
\newblock Microstorms in cellular polymers: A route to soft piezoelectric
  transducer materials with engineered macroscopic dipoles.
\newblock \emph{ChemPhysChem}, 6\penalty0 (6):\penalty0 1014--1025, 2005.

\bibitem[Buchberger et~al.(2008)Buchberger, Schw{\"o}diauer, and
  Bauer]{buchberger2008flexible}
Gerda Buchberger, Reinhard Schw{\"o}diauer, and Siegfried Bauer.
\newblock Flexible large area ferroelectret sensors for location sensitive
  touchpads.
\newblock \emph{Applied Physics Letters}, 92\penalty0 (12):\penalty0 123511,
  2008.

\bibitem[Deng et~al.(2014{\natexlab{a}})Deng, Liu, and
  Sharma]{deng2014electrets}
Qian Deng, Liping Liu, and Pradeep Sharma.
\newblock Electrets in soft materials: Nonlinearity, size effects, and giant
  electromechanical coupling.
\newblock \emph{Physical Review E}, 90\penalty0 (1):\penalty0 012603,
  2014{\natexlab{a}}.

\bibitem[Liu and Sharma(2018)]{liu2018emergent}
Liping Liu and Pradeep Sharma.
\newblock Emergent electromechanical coupling of electrets and some exact
  relations—the effective properties of soft materials with embedded external
  charges and dipoles.
\newblock \emph{Journal of the Mechanics and Physics of Solids}, 112:\penalty0
  1--24, 2018.

\bibitem[Darbaniyan et~al.(2019)Darbaniyan, Dayal, Liu, and
  Sharma]{darbaniyan2019designing}
Faezeh Darbaniyan, Kaushik Dayal, Liping Liu, and Pradeep Sharma.
\newblock Designing soft pyroelectric and electrocaloric materials using
  electrets.
\newblock \emph{Soft matter}, 15\penalty0 (2):\penalty0 262--277, 2019.

\bibitem[Darbaniyan et~al.(2021)Darbaniyan, Mozaffari, Liu, and
  Sharma]{darbaniyan2021soft}
Faezeh Darbaniyan, Kosar Mozaffari, Liping Liu, and Pradeep Sharma.
\newblock Soft matter mechanics and the mechanisms underpinning the infrared
  vision of snakes.
\newblock \emph{Matter}, 4\penalty0 (1):\penalty0 241--252, 2021.

\bibitem[Apte et~al.(2020)Apte, Mozaffari, Samghabadi, Hachtel, Chang, Susarla,
  Idrobo, Moore, Glavin, Litvinov, et~al.]{apte20202d}
Amey Apte, Kosar Mozaffari, Farnaz~Safi Samghabadi, Jordan~A Hachtel, Long
  Chang, Sandhya Susarla, Juan~Carlos Idrobo, David~C Moore, Nicholas~R Glavin,
  Dmitri Litvinov, et~al.
\newblock 2d electrets of ultrathin moo2 with apparent piezoelectricity.
\newblock \emph{Advanced Materials}, 32\penalty0 (24):\penalty0 2000006, 2020.

\bibitem[Mellinger et~al.(2006)Mellinger, Wegener, Wirges, Mallepally, and
  Gerhard-Multhaupt]{mellinger2006thermal}
Axel Mellinger, Michael Wegener, Werner Wirges, Rajendar~Reddy Mallepally, and
  Reimund Gerhard-Multhaupt.
\newblock Thermal and temporal stability of ferroelectret films made from
  cellular polypropylene/air composites.
\newblock \emph{Ferroelectrics}, 331\penalty0 (1):\penalty0 189--199, 2006.

\bibitem[Krichen and Sharma(2016)]{krichen2016flexoelectricity}
Sana Krichen and Pradeep Sharma.
\newblock Flexoelectricity: A perspective on an unusual electromechanical
  coupling.
\newblock \emph{Journal of Applied Mechanics}, 83\penalty0 (3), 2016.

\bibitem[Deng et~al.(2017)Deng, Liu, and Sharma]{deng2017continuum}
Qian Deng, Liping Liu, and Pradeep Sharma.
\newblock A continuum theory of flexoelectricity.
\newblock In \emph{Flexoelectricity in Solids: From Theory to Applications},
  pages 111--167. World Scientific, 2017.

\bibitem[Nguyen et~al.(2013)Nguyen, Mao, Yeh, Purohit, and
  McAlpine]{nguyen2013nanoscale}
Thanh~D Nguyen, Sheng Mao, Yao-Wen Yeh, Prashant~K Purohit, and Michael~C
  McAlpine.
\newblock Nanoscale flexoelectricity.
\newblock \emph{Advanced Materials}, 25\penalty0 (7):\penalty0 946--974, 2013.

\bibitem[Sharma et~al.(2010)Sharma, Landis, and
  Sharma]{sharma2010piezoelectric}
ND~Sharma, CM~Landis, and P~Sharma.
\newblock Piezoelectric thin-film superlattices without using piezoelectric
  materials.
\newblock \emph{Journal of Applied Physics}, 108\penalty0 (2):\penalty0 024304,
  2010.

\bibitem[Fousek et~al.(1999)Fousek, Cross, and Litvin]{fousek1999possible}
J~Fousek, LE~Cross, and DB~Litvin.
\newblock Possible piezoelectric composites based on the flexoelectric effect.
\newblock \emph{Materials Letters}, 39\penalty0 (5):\penalty0 287--291, 1999.

\bibitem[Chu et~al.(2009)Chu, Zhu, Li, and Cross]{chu2009flexure}
Baojin Chu, Wenyi Zhu, Nan Li, and L~Eric Cross.
\newblock Flexure mode flexoelectric piezoelectric composites.
\newblock \emph{Journal of Applied Physics}, 106\penalty0 (10):\penalty0
  104109, 2009.

\bibitem[Wang et~al.(2019)Wang, Yang, and Sharma]{wang2019flexoelectricity}
Binglei Wang, Shengyou Yang, and Pradeep Sharma.
\newblock Flexoelectricity as a universal mechanism for energy harvesting from
  crumpling of thin sheets.
\newblock \emph{Phys. Rev. B}, 100:\penalty0 035438, Jul 2019.
\newblock \doi{10.1103/PhysRevB.100.035438}.
\newblock URL \url{https://link.aps.org/doi/10.1103/PhysRevB.100.035438}.

\bibitem[Deng et~al.(2014{\natexlab{b}})Deng, Kammoun, Erturk, and
  Sharma]{deng2014nanoscale}
Qian Deng, Mejdi Kammoun, Alper Erturk, and Pradeep Sharma.
\newblock Nanoscale flexoelectric energy harvesting.
\newblock \emph{International Journal of Solids and Structures}, 51\penalty0
  (18):\penalty0 3218--3225, 2014{\natexlab{b}}.

\bibitem[Jiang et~al.(2013)Jiang, Huang, and Zhang]{jiang2013flexoelectric}
Xiaoning Jiang, Wenbin Huang, and Shujun Zhang.
\newblock Flexoelectric nano-generator: Materials, structures and devices.
\newblock \emph{Nano Energy}, 2\penalty0 (6):\penalty0 1079--1092, 2013.

\bibitem[Majdoub et~al.(2008)Majdoub, Sharma, and
  {\c{C}}a{\u{g}}in]{majdoub2008dramatic}
MS~Majdoub, P~Sharma, and T~{\c{C}}a{\u{g}}in.
\newblock Dramatic enhancement in energy harvesting for a narrow range of
  dimensions in piezoelectric nanostructures.
\newblock \emph{Physical Review B}, 78\penalty0 (12):\penalty0 121407, 2008.

\bibitem[Majdoub et~al.(2009)Majdoub, Sharma, and
  {\c{C}}a{\u{g}}in]{majdoub2009erratum}
MS~Majdoub, P~Sharma, and T~{\c{C}}a{\u{g}}in.
\newblock Erratum: Dramatic enhancement in energy harvesting for a narrow range
  of dimensions in piezoelectric nanostructures [phys. rev. b 78, 121407
  (r)(2008)].
\newblock \emph{Physical Review B}, 79\penalty0 (15):\penalty0 159901, 2009.

\bibitem[Rahmati et~al.(2019)Rahmati, Bauer, and Sharma]{rahmati2019nonlinear}
Amir~Hossein Rahmati, Siegfried Bauer, and Pradeep Sharma.
\newblock Nonlinear bending deformation of soft electrets and prospects for
  engineering flexoelectricity and transverse (d 31) piezoelectricity.
\newblock \emph{Soft matter}, 15\penalty0 (1):\penalty0 127--148, 2019.

\bibitem[Choi and Kim(2017)]{choi2017measurement}
Seung-Bok Choi and Gi-Woo Kim.
\newblock Measurement of flexoelectric response in polyvinylidene fluoride
  films for piezoelectric vibration energy harvesters.
\newblock \emph{Journal of Physics D: Applied Physics}, 50\penalty0
  (7):\penalty0 075502, 2017.

\bibitem[Bhaskar et~al.(2016{\natexlab{a}})Bhaskar, Banerjee, Abdollahi,
  Solanas, Rijnders, and Catalan]{bhaskar2016flexoelectric}
Umesh~Kumar Bhaskar, Nirupam Banerjee, Amir Abdollahi, E~Solanas, Guus
  Rijnders, and Gustau Catalan.
\newblock Flexoelectric mems: towards an electromechanical strain diode.
\newblock \emph{Nanoscale}, 8\penalty0 (3):\penalty0 1293--1298,
  2016{\natexlab{a}}.

\bibitem[Bhaskar et~al.(2016{\natexlab{b}})Bhaskar, Banerjee, Abdollahi, Wang,
  Schlom, Rijnders, and Catalan]{bhaskar2016bflexoelectric}
Umesh~Kumar Bhaskar, Nirupam Banerjee, Amir Abdollahi, Zhe Wang, Darrell~G
  Schlom, Guus Rijnders, and Gustau Catalan.
\newblock A flexoelectric microelectromechanical system on silicon.
\newblock \emph{Nature nanotechnology}, 11\penalty0 (3):\penalty0 263--266,
  2016{\natexlab{b}}.

\bibitem[Wang et~al.(2013)Wang, Zhang, Wang, Yue, Li, Miao, and
  Zhu]{wang2013giant}
Zhihong Wang, Xi~Xiang Zhang, Xianbin Wang, Weisheng Yue, Jingqi Li, Jianmin
  Miao, and Weiguang Zhu.
\newblock Giant flexoelectric polarization in a micromachined ferroelectric
  diaphragm.
\newblock \emph{Advanced Functional Materials}, 23\penalty0 (1):\penalty0
  124--132, 2013.

\bibitem[Abdollahi and Arias(2015)]{abdollahi2015constructive}
Amir Abdollahi and Irene Arias.
\newblock Constructive and destructive interplay between piezoelectricity and
  flexoelectricity in flexural sensors and actuators.
\newblock \emph{Journal of Applied Mechanics}, 82\penalty0 (12), 2015.

\bibitem[Deng et~al.(2019)Deng, Ahmadpoor, Brownell, and
  Sharma]{deng2019collusion}
Qian Deng, Fatemeh Ahmadpoor, William~E Brownell, and Pradeep Sharma.
\newblock The collusion of flexoelectricity and hopf bifurcation in the hearing
  mechanism.
\newblock \emph{Journal of the Mechanics and Physics of Solids}, 130:\penalty0
  245--261, 2019.

\bibitem[Brownell et~al.(2001)Brownell, Spector, Raphael, and
  Popel]{brownell2001micro}
WE~Brownell, AA~Spector, RM~Raphael, and Aleksander~S Popel.
\newblock Micro-and nanomechanics of the cochlear outer hair cell.
\newblock \emph{Annual review of biomedical engineering}, 3\penalty0
  (1):\penalty0 169--194, 2001.

\bibitem[Breneman et~al.(2009)Breneman, Brownell, and
  Rabbitt]{breneman2009hair}
Kathryn~D Breneman, William~E Brownell, and Richard~D Rabbitt.
\newblock Hair cell bundles: flexoelectric motors of the inner ear.
\newblock \emph{PLoS One}, 4\penalty0 (4):\penalty0 e5201, 2009.

\bibitem[Deng et~al.(2014{\natexlab{c}})Deng, Liu, and
  Sharma]{deng2014flexoelectricity}
Qian Deng, Liping Liu, and Pradeep Sharma.
\newblock Flexoelectricity in soft materials and biological membranes.
\newblock \emph{Journal of the Mechanics and Physics of Solids}, 62:\penalty0
  209--227, 2014{\natexlab{c}}.

\bibitem[Wen et~al.(2019)Wen, Li, Tan, Deng, and Shen]{wen2019flexoelectret}
Xin Wen, Dongfan Li, Kai Tan, Qian Deng, and Shengping Shen.
\newblock Flexoelectret: an electret with a tunable flexoelectriclike response.
\newblock \emph{Physical review letters}, 122\penalty0 (14):\penalty0 148001,
  2019.

\bibitem[Maranganti et~al.(2006)Maranganti, Sharma, and
  Sharma]{maranganti2006electromechanical}
R~Maranganti, ND~Sharma, and P~Sharma.
\newblock Electromechanical coupling in nonpiezoelectric materials due to
  nanoscale nonlocal size effects: Green’s function solutions and embedded
  inclusions.
\newblock \emph{Physical Review B}, 74\penalty0 (1):\penalty0 014110, 2006.

\bibitem[Codony et~al.(2021)Codony, Gupta, Marco, and
  Arias]{codony2021modeling}
David Codony, Prakhar Gupta, Onofre Marco, and Irene Arias.
\newblock Modeling flexoelectricity in soft dielectrics at finite deformation.
\newblock \emph{Journal of the Mechanics and Physics of Solids}, 146:\penalty0
  104182, 2021.

\bibitem[Abdollahi et~al.(2014)Abdollahi, Peco, Millan, Arroyo, and
  Arias]{abdollahi2014computational}
Amir Abdollahi, Christian Peco, Daniel Millan, Marino Arroyo, and Irene Arias.
\newblock Computational evaluation of the flexoelectric effect in dielectric
  solids.
\newblock \emph{Journal of Applied Physics}, 116\penalty0 (9):\penalty0 093502,
  2014.

\bibitem[Yvonnet and Liu(2017)]{yvonnet2017numerical}
Julien Yvonnet and LP~Liu.
\newblock A numerical framework for modeling flexoelectricity and maxwell
  stress in soft dielectrics at finite strains.
\newblock \emph{Computer Methods in Applied Mechanics and Engineering},
  313:\penalty0 450--482, 2017.

\bibitem[Thai et~al.(2021)Thai, Zhuang, Park, and Rabczuk]{thai2021staggered}
Tran~Quoc Thai, Xiaoying Zhuang, Harold~S Park, and Timon Rabczuk.
\newblock A staggered explicit-implicit isogeometric formulation for large
  deformation flexoelectricity.
\newblock \emph{Engineering Analysis with Boundary Elements}, 122:\penalty0
  1--12, 2021.

\bibitem[Tagantsev(1986)]{tagantsev1986piezoelectricity}
AK~Tagantsev.
\newblock Piezoelectricity and flexoelectricity in crystalline dielectrics.
\newblock \emph{Physical Review B}, 34\penalty0 (8):\penalty0 5883, 1986.

\bibitem[Maranganti and Sharma(2009)]{maranganti2009atomistic}
R~Maranganti and P~Sharma.
\newblock Atomistic determination of flexoelectric properties of crystalline
  dielectrics.
\newblock \emph{Physical Review B}, 80\penalty0 (5):\penalty0 054109, 2009.

\bibitem[Stengel(2013)]{stengel2013flexoelectricity}
Massimiliano Stengel.
\newblock Flexoelectricity from density-functional perturbation theory.
\newblock \emph{Physical Review B}, 88\penalty0 (17):\penalty0 174106, 2013.

\bibitem[Stengel(2014)]{stengel2014surface}
Massimiliano Stengel.
\newblock Surface control of flexoelectricity.
\newblock \emph{Physical Review B}, 90\penalty0 (20):\penalty0 201112, 2014.

\bibitem[Dreyer et~al.(2018)Dreyer, Stengel, and Vanderbilt]{dreyer2018current}
Cyrus~E Dreyer, Massimiliano Stengel, and David Vanderbilt.
\newblock Current-density implementation for calculating flexoelectric
  coefficients.
\newblock \emph{Physical Review B}, 98\penalty0 (7):\penalty0 075153, 2018.

\bibitem[deGennes and Prost(1993)]{de1993physics}
Pierre-Gilles deGennes and Jacques Prost.
\newblock \emph{The physics of liquid crystals}, volume~83.
\newblock Oxford university press, 1993.

\bibitem[Meyer(1969)]{meyer1969piezoelectric}
Robert~B Meyer.
\newblock Piezoelectric effects in liquid crystals.
\newblock \emph{Physical Review Letters}, 22\penalty0 (18):\penalty0 918, 1969.

\bibitem[Kothari et~al.(2018)Kothari, Cha, and Kim]{kothari2018critical}
Mrityunjay Kothari, Moon-Hyun Cha, and Kyung-Suk Kim.
\newblock Critical curvature localization in graphene. i.
  quantum-flexoelectricity effect.
\newblock \emph{Proceedings of the Royal Society A: Mathematical, Physical and
  Engineering Sciences}, 474\penalty0 (2214):\penalty0 20180054, 2018.

\bibitem[Kothari et~al.(2019)Kothari, Cha, Lefevre, and
  Kim]{kothari2019critical}
Mrityunjay Kothari, Moon-Hyun Cha, Victor Lefevre, and Kyung-Suk Kim.
\newblock Critical curvature localization in graphene. ii. non-local
  flexoelectricity--dielectricity coupling.
\newblock \emph{Proceedings of the Royal Society A}, 475\penalty0
  (2221):\penalty0 20180671, 2019.

\bibitem[Kvashnin et~al.(2015)Kvashnin, Sorokin, and
  Yakobson]{kvashnin2015flexoelectricity}
Alexander~G Kvashnin, Pavel~B Sorokin, and Boris~I Yakobson.
\newblock Flexoelectricity in carbon nanostructures: nanotubes, fullerenes, and
  nanocones.
\newblock \emph{The journal of physical chemistry letters}, 6\penalty0
  (14):\penalty0 2740--2744, 2015.

\bibitem[Ahmadpoor and Sharma(2015)]{ahmadpoor2015flexoelectricity}
Fatemeh Ahmadpoor and Pradeep Sharma.
\newblock Flexoelectricity in two-dimensional crystalline and biological
  membranes.
\newblock \emph{Nanoscale}, 7\penalty0 (40):\penalty0 16555--16570, 2015.

\bibitem[Kumar et~al.(2021)Kumar, Codony, Arias, and
  Suryanarayana]{kumar2021flexoelectricity}
Shashikant Kumar, David Codony, Irene Arias, and Phanish Suryanarayana.
\newblock Flexoelectricity in atomic monolayers from first principles.
\newblock \emph{Nanoscale}, 2021.

\bibitem[Banerjee and Suryanarayana(2016)]{banerjee2016cyclic}
Amartya~S Banerjee and Phanish Suryanarayana.
\newblock Cyclic density functional theory: A route to the first principles
  simulation of bending in nanostructures.
\newblock \emph{Journal of the Mechanics and Physics of Solids}, 96:\penalty0
  605--631, 2016.

\bibitem[Marvan and Havr{\'a}nek(1997)]{marvan1997static}
M~Marvan and A~Havr{\'a}nek.
\newblock Static volume flexoelectric effect in a model of linear chains.
\newblock \emph{Solid state communications}, 101\penalty0 (7):\penalty0
  493--496, 1997.

\bibitem[Marvan and Havr{\'a}nek(1998)]{marvan1998flexoelectric}
M~Marvan and A~Havr{\'a}nek.
\newblock Flexoelectric effect in elastomers.
\newblock In \emph{Relationships of Polymeric Structure and Properties}, pages
  33--36. Springer, 1998.

\bibitem[Baskaran et~al.(2012)Baskaran, He, Wang, and Fu]{baskaran2012strain}
Sivapalan Baskaran, Xiangtong He, Yu~Wang, and John~Y Fu.
\newblock Strain gradient induced electric polarization in $\alpha$-phase
  polyvinylidene fluoride films under bending conditions.
\newblock \emph{Journal of Applied Physics}, 111\penalty0 (1):\penalty0 014109,
  2012.

\bibitem[Chu and Salem(2012)]{chu2012flexoelectricity}
Baojin Chu and DR~Salem.
\newblock Flexoelectricity in several thermoplastic and thermosetting polymers.
\newblock \emph{Applied Physics Letters}, 101\penalty0 (10):\penalty0 103905,
  2012.

\bibitem[Treloar(1975)]{treloar1975physics}
L~R~G Treloar.
\newblock \emph{The physics of rubber elasticity}.
\newblock Oxford University Press, 1975.

\bibitem[Flory(1944)]{flory1944network}
Paul~J Flory.
\newblock Network structure and the elastic properties of vulcanized rubber.
\newblock \emph{Chemical reviews}, 35\penalty0 (1):\penalty0 51--75, 1944.

\bibitem[Arruda and Boyce(1993)]{arruda1993threee}
Ellen~M Arruda and Mary~C Boyce.
\newblock A three-dimensional constitutive model for the large stretch behavior
  of rubber elastic materials.
\newblock \emph{Journal of the Mechanics and Physics of Solids}, 41\penalty0
  (2):\penalty0 389--412, 1993.

\bibitem[Boyce and Arruda(2000)]{boyce2000constitutive}
Mary~C Boyce and Ellen~M Arruda.
\newblock Constitutive models of rubber elasticity: a review.
\newblock \emph{Rubber chemistry and technology}, 73\penalty0 (3):\penalty0
  504--523, 2000.

\bibitem[Cohen and deBotton(2016)]{cohen2016electromechanical}
Noy Cohen and Gal deBotton.
\newblock Electromechanical interplay in deformable dielectric elastomer
  networks.
\newblock \emph{Physical review letters}, 116\penalty0 (20):\penalty0 208303,
  2016.

\bibitem[Itskov et~al.(2018)Itskov, Khi{\^e}m, and
  Waluyo]{itskov2018electroelasticity}
Mikhail Itskov, Vu~Ngoc Khi{\^e}m, and Sugeng Waluyo.
\newblock Electroelasticity of dielectric elastomers based on molecular chain
  statistics.
\newblock \emph{Mathematics and Mechanics of Solids}, 2018.

\bibitem[Grasinger and Dayal(2020{\natexlab{a}})]{grasinger2020architected}
Matthew Grasinger and Kaushik Dayal.
\newblock Architected elastomer networks for optimal electromechanical
  response.
\newblock \emph{Journal of the Mechanics and Physics of Solids}, page 104171,
  2020{\natexlab{a}}.
\newblock ISSN 0022-5096.

\bibitem[Grasinger and Dayal(2020{\natexlab{b}})]{grasinger2020statistical}
Matthew Grasinger and Kaushik Dayal.
\newblock Statistical mechanical analysis of the electromechanical coupling in
  an electrically-responsive polymer chain.
\newblock \emph{Soft Matter}, 16:\penalty0 6265--6284, 2020{\natexlab{b}}.

\bibitem[Kuhn and Gr{\"u}n(1942)]{kuhn1942beziehungen}
Werner Kuhn and F~Gr{\"u}n.
\newblock Beziehungen zwischen elastischen konstanten und
  dehnungsdoppelbrechung hochelastischer stoffe.
\newblock \emph{Kolloid-Zeitschrift}, 101\penalty0 (3):\penalty0 248--271,
  1942.

\bibitem[Lighthill(1958)]{lighthill1958introduction}
Michael~J Lighthill.
\newblock \emph{An introduction to Fourier analysis and generalised functions}.
\newblock Cambridge University Press, 1958.

\bibitem[Fredrickson et~al.(2006)]{fredrickson2006equilibrium}
Glenn Fredrickson et~al.
\newblock \emph{The equilibrium theory of inhomogeneous polymers}, volume 134.
\newblock Oxford University Press on Demand, 2006.

\bibitem[Lide(2004)]{lide2004crc}
David~R Lide.
\newblock \emph{CRC handbook of chemistry and physics}, volume~85.
\newblock CRC press, 2004.

\bibitem[Tadmor and Miller(2011)]{tadmor2011modeling}
Ellad~B Tadmor and Ronald~E Miller.
\newblock \emph{Modeling materials: continuum, atomistic and multiscale
  techniques}.
\newblock Cambridge University Press, 2011.

\bibitem[James and Guth(1943)]{james1943theory}
Hubert~M James and Eugene Guth.
\newblock Theory of the elastic properties of rubber.
\newblock \emph{The Journal of Chemical Physics}, 11\penalty0 (10):\penalty0
  455--481, 1943.

\bibitem[Flory and Rehner(1943)]{flory1943jr}
Paul~J Flory and John Rehner.
\newblock Statistical mechanics of cross-linked polymer networks. ii.
\newblock \emph{The Journal of Chemical Physics}, 11:\penalty0 521--526, 1943.

\bibitem[Wu and Van Der~Giessen(1993)]{wu1993improved}
PD~Wu and Erik Van Der~Giessen.
\newblock On improved network models for rubber elasticity and their
  applications to orientation hardening in glassy polymers.
\newblock \emph{Journal of the Mechanics and Physics of Solids}, 41\penalty0
  (3):\penalty0 427--456, 1993.

\bibitem[Cohen et~al.(2016)Cohen, Dayal, and
  deBotton]{cohen2016electroelasticity}
Noy Cohen, Kaushik Dayal, and Gal deBotton.
\newblock Electroelasticity of polymer networks.
\newblock \emph{Journal of Mechanics Physics of Solids}, 92:\penalty0 105--126,
  2016.

\bibitem[Grasinger(2019)]{grasinger2019multiscale}
Matthew Grasinger.
\newblock \emph{Multiscale Modeling and Theoretical Design of Dielectric
  Elastomers}.
\newblock PhD thesis, Carnegie Mellon University, 2019.

\bibitem[Grasinger et~al.()Grasinger, Majidi, and
  Dayal]{grasingerIPnongaussian}
Matthew Grasinger, Carmel Majidi, and Kaushik Dayal.
\newblock Nonlinear statistical mechanics drives intrinsic electrostriction and
  volumetric torque in polymer networks.
\newblock \emph{In preparation}.

\bibitem[Baskaran et~al.(2011)Baskaran, He, Chen, and
  Fu]{baskaran2011experimental}
Sivapalan Baskaran, Xiangtong He, Qin Chen, and John~Y Fu.
\newblock Experimental studies on the direct flexoelectric effect in
  $\alpha$-phase polyvinylidene fluoride films.
\newblock \emph{Applied Physics Letters}, 98\penalty0 (24):\penalty0 242901,
  2011.

\bibitem[Fukada and Takashita(1969)]{fukada1969piezoelectric}
Eiichi Fukada and Saburo Takashita.
\newblock Piezoelectric effect in polarized poly (vinylidene fluoride).
\newblock \emph{Japanese Journal of Applied Physics}, 8\penalty0 (7):\penalty0
  960, 1969.

\bibitem[Sappati and Bhadra(2018)]{sappati2018piezoelectric}
Kiran~Kumar Sappati and Sharmistha Bhadra.
\newblock Piezoelectric polymer and paper substrates: a review.
\newblock \emph{Sensors}, 18\penalty0 (11):\penalty0 3605, 2018.

\bibitem[Ashraf et~al.(2009)Ashraf, Millare, Gerasimenko, Bao, Pandey, Lake,
  and Vullev]{ashraf2009theoretical}
MK~Ashraf, Brent Millare, Alexander~A Gerasimenko, Duoduo Bao, Rajeev~R Pandey,
  Roger~K Lake, and Valentine~I Vullev.
\newblock Theoretical design of bioinspired macromolecular electrets based on
  anthranilamide derivatives.
\newblock \emph{Biotechnology progress}, 25\penalty0 (4):\penalty0 915--922,
  2009.

\bibitem[Griffiths(2017)]{griffiths2005introduction}
David~J Griffiths.
\newblock Introduction to electrodynamics, 2017.

\end{thebibliography}

\end{document}